\documentclass[letterpaper,journal]{IEEEtran}
\usepackage{amsmath,amsfonts,amssymb,mathtools,bm}
\usepackage{array,multirow}
\usepackage[caption=false,font=normalsize,labelfont=sf,textfont=sf]{subfig}
\usepackage{stfloats}
\usepackage{url,graphicx,cite}
\usepackage{amsthm}
\usepackage{xcolor}
\usepackage{hyperref}
\hypersetup{colorlinks=true,citecolor=blue,linkcolor=blue,urlcolor=blue}
\newtheorem{theorem}{Theorem}
\newtheorem{lemma}{Lemma}
\newtheorem{corollary}{Corollary}

\renewcommand{\mathbf}[1]{\bm{#1}}

\newcommand{\ms}[2]{\shortstack{$#1$\\[-1pt]{\footnotesize $\pm\,#2$}}}

\begin{document}

\title{Quantum Orthogonal Separable Physics-Informed Neural Networks}

\author{
    \IEEEauthorblockN{
        Pietro Zanotta\IEEEauthorrefmark{1},
        Ljubomir Budinski\IEEEauthorrefmark{2}\textsuperscript{,}\IEEEauthorrefmark{3},
        Çağlar Aytekin\IEEEauthorrefmark{2},
        Valtteri Lahtinen\IEEEauthorrefmark{2}\textsuperscript{,}\IEEEauthorrefmark{4}
    }
    \\[0.5ex]
    \IEEEauthorblockA{
        \IEEEauthorrefmark{1}Johns Hopkins University, Baltimore, USA \quad
        \IEEEauthorrefmark{2}Quanscient Oy, Tampere, Finland \\
        \IEEEauthorrefmark{3}University of Novi Sad, Novi Sad, Serbia \\
        \IEEEauthorrefmark{4}Lappeenranta--Lahti University of Technology, Lappeenranta, Finland \\[0.5ex]
        Email: pzanott1@jh.edu, \{ljubomir.budinski, caglar.aytekin, valtteri.lahtinen\}@quanscient.com
    }
    \thanks{This work was done during P.Z.'s internship at Quanscient Oy. V.L. acknowledges support from the Research Council of Finland (Flagship of Advanced Mathematics for Sensing Imaging and Modelling, grant 359183).}
    \thanks{Corresponding author: Pietro Zanotta (pzanott1@jh.edu).}
}

\IEEEpubid{This work has been submitted to the IEEE for possible publication. Copyright may be transferred without notice, after which this version may no longer be accessible.}

\markboth{}{Zanotta \MakeLowercase{\textit{et al.}}: Quantum Orthogonal Separable Physics-Informed Neural Networks}
\maketitle

\IEEEpubid{This work has been submitted to the IEEE for possible publication. Copyright may be transferred without notice, after which this version may no longer be accessible.}

\begin{abstract}
We introduce Quantum Orthogonal Separable Physics-Informed Neural Networks (QO-SPINNs), which replace selected dense hidden transformations in physics-informed PDE models with quantum-compatible orthogonal maps. A pyramidal reconfigurable-beamsplitter circuit realizes these maps with $\Theta(d)$ circuit depth. Under the stated signed-readout access model and fixed per-call failure probability, a width-$d$ map has sequential access-cost upper bound $\mathcal O(d\log d/\epsilon^2)$, which is asymptotically smaller than the $\Theta(d^2)$ cost of conventional direct dense multiplication when $\epsilon(d)^{-2}\log d=o(d)$. Because unary state preparation discards magnitude, we retain each hidden-state norm classically and restore it after readout, yielding the exact map $Wh$ under ideal readout. We study this construction in two complementary settings: forward PDE regression and a norm-controlled residual model for representation geometry and uncertainty quantification. In the norm-controlled model, structural orthogonality yields explicit bi-Lipschitz bounds, while finite-readout analysis relates sufficient reconstruction precision to a prescribed physical-coordinate resolution through trained-model constants. This makes the access-cost condition assessable for trained checkpoints, although favorable fixed-resolution scaling across widths is not established. A post-hoc Bayesian-linear covariance provides uncertainty from the same frozen predictor, and cross-PDE transfer shows that strong distance-awareness can persist even when absolute coverage calibration fails. All Quantum Orthogonal (QO) layers are evaluated by classical simulation: the results therefore characterize a quantum-compatible structured neural primitive rather than demonstrating a practical quantum-hardware advantage.
\end{abstract}


\section{Introduction}
\label{sec:introduction}

Physics-informed neural networks (PINNs) approximate PDE solutions by combining available initial, boundary, and observational data with residuals of the governing equations (Ref.~\cite{raissi2019physics,karniadakis2021physics}). Recent work has advanced this framework along several complementary directions. Uncertainty evaluation has been used to adaptively weight physics-informed losses for robust prediction (Ref.~\cite{hua2024weighted}), latent-space dynamics have been introduced to improve generalization across PDE configurations (Ref.~\cite{wang2025pido}), and physics-informed priors have been incorporated into Bayesian neural networks for uncertainty quantification (Ref.~\cite{menicali2025pibnn}); related work has also developed a posteriori error certification (Ref.~\cite{hillebrecht2025error}). Beyond residual-based PINNs, recent work has also incorporated geometric structure into learned physical models to reproduce conservation and dissipation laws (Ref.~\cite{matsubara2025deep}), while a recent survey provides a broader account of deep neural approaches for PDEs (Ref.~\cite{huang2025pdesurvey}). Separable PINNs (SPINNs) address computational cost by evaluating coordinate-wise subnetworks and combining their outputs through a separable representation (Ref.~\cite{cho2023separable}), with recent extensions to high-dimensional kinetic equations (Ref.~\cite{oh2025separable}). These developments target training robustness, cross-configuration generalization, uncertainty modeling, certification, physics-preserving dynamics, or separable evaluation. QO-SPINN instead focuses on the structure and execution of the square hidden transformations inside the coordinate subnetworks, replacing selected dense maps by norm-restored orthogonal transformations and analyzing their geometric and readout properties.

\IEEEpubidadjcol

Orthogonal hidden maps offer a natural structured alternative. Pyramidal RBS circuits parameterize $W\in SO(d)$ exactly and can be evaluated classically in $\mathcal O(d^2)$ operations or implemented in the unary quantum subspace with $\Theta(d)$ circuit depth as shown in Ref.~\cite{kerenidis2021classical,landman2022quantum}. Recent quantum scientific-learning approaches include quantum neural operators (Ref.~\cite{xiao2025quantum}) and quantum physics-informed neural networks in which parameterized quantum circuits participate directly in the PDE solution architecture (Ref.~\cite{berger2025trainable,panichi2026quantum}). QO-SPINN addresses a different computational question: the physics-informed architecture retains classical nonlinear processing, while the quantum-compatible circuit realizes selected learned orthogonal hidden transformations. Consequently, the computational claim studied here concerns execution and signed readout of those hidden maps rather than an end-to-end quantum PDE solver. Trainability, encoding, and measurement nevertheless remain central considerations as studied in Ref.~\cite{monbroussou2025trainability,coyle2025density,rahman2026qcnn}, and readout cannot be treated as free: under the signed $\ell_\infty$ tomography model used here, the execution cost depends explicitly on the requested reconstruction tolerance.

There is, however, a more immediate difficulty in inserting the unary construction into a neural network. Amplitude encoding represents the direction $h/\|h\|_2$ of a hidden state $h\in\mathbb R^d$, but not its magnitude $\|h\|_2$. A quantum implementation that transforms only the normalized state therefore does not by itself reproduce the corresponding classical hidden map. To preserve that map, the radial information must also cross the quantum interface. This motivates the norm-restored orthogonal layer at the core of QO-SPINN.

QO-SPINN is built around a simple resolution of this problem. For $h\in\mathbb R^d$, write $h=\|h\|_2q(h)$ with $q(h)=~h/\|h\|_2$ for $h\neq0$ and $q(0)=0$. The circuit acts on the direction $q(h)$, while $\|h\|_2$ is retained classically and restored after readout. Under exact readout, the resulting map is $\mathcal R_W(h):=~\|h\|_2Wq(h)=Wh$, so unary state preparation changes the execution of the hidden transformation without changing the transformation itself. We use this primitive in a norm-conditioned FiLM SPINN for forward regression and in a norm-controlled residual model for representation geometry and uncertainty quantification.

This construction is useful for more than its quantum execution path. Because every Quantum Orthogonal (QO) transformation has to satisfy $\|W\|_2=~1$ to be a valid quantum circuit, the norm-controlled residual architecture admits explicit two-sided bounds on latent distances and, consequently, on the induced RBF kernel. Finite directional-readout error can then be propagated through the same representation, yielding a sufficient relation between tomography accuracy $\epsilon$ and the physical-coordinate scale that remains distinguishable after readout. A dense residual network with exact spectral normalization satisfies the same operator-norm hypothesis; the difference is that QO enforces unit norm structurally rather than through a separately computed normalization factor.

The same norm-controlled representation also provides a natural basis for uncertainty quantification. After deterministic physics-informed training, we construct a post-hoc Bayesian-linear covariance from the frozen features while retaining the learned predictive coefficients. This makes it possible to study latent geometry and uncertainty scale separately, including under transfer from the Heat equation to Burgers. In parallel, the QO-FiLM experiments address the complementary question of whether the orthogonal primitive remains effective for forward PDE regression.
These ideas lead to three main contributions:
\begin{itemize}
    \item \textbf{Norm-restored quantum orthogonal layers for separable physics-informed learning.} We introduce a QO-SPINN primitive that retains the norm removed by unary state preparation and restores it after the pyramidal orthogonal transformation, so ideal execution exactly recovers the intended map $Wh$. We embed this primitive in architectures for forward PDE regression and uncertainty-aware representation learning.

    \item \textbf{A computational and geometric analysis of the QO primitive.} Under the stated signed-readout access model, we identify the condition $\epsilon(d)^{-2}\log d=o(d)$, at fixed per-call failure probability, under which the per-map sequential access-cost upper bound is asymptotically smaller than conventional direct dense multiplication. For the norm-controlled residual architecture, structural orthogonality yields explicit bi-Lipschitz and latent-kernel bounds, while finite readout gives an additive distance-distortion bound and a model-dependent sufficient precision for preserving prescribed physical-coordinate resolution. These trained-model quantities allow the fixed-resolution access-cost condition to be assessed without assuming width-independent precision.

    \item \textbf{Matched empirical evaluation of prediction, norm control, and uncertainty.} Across forward PDE benchmarks we evaluate the predictive viability of QO layers against matched dense transformations. In the norm-controlled setting we validate the geometric and finite-readout bounds, compare structural orthogonality with exact and approximate dense spectral normalization, and study post-hoc uncertainty under covariance-hyperparameter transfer from Heat to Burgers.
\end{itemize}

\section{Quantum Orthogonal Layers and Readout Model}
\label{sec:computational_setting}

We begin with the hidden transformation that QO-SPINN replaces. After recalling the separable PINN representation, we describe the pyramidal RBS parametrization of an orthogonal map and the signed-readout model needed when nonlinearities are evaluated classically between quantum layers. This gives the execution model used later to compare a width-$d$ QO transformation with direct dense multiplication.

\paragraph{\textbf{Separable Physics-Informed Approximation}}
\label{sec:separable_setting}

Let $\mathbf{x}=(x_1,\ldots,x_K)\in\Omega\subset\mathbb{R}^K$ denote the physical coordinates and let $u_\theta(\mathbf{x})$ be a differentiable neural approximation of the PDE solution. We use the standard physics-informed objective
\begin{equation}
\mathcal{L}(\theta)=\lambda_{\rm obs}\mathcal{L}_{\rm obs}+\lambda_{\rm bc}\mathcal{L}_{\rm bc}+\lambda_{\rm pde}\mathcal{L}_{\rm pde},
\label{eq:generic_pinn_loss}
\end{equation}
where the individual terms enforce available observations, initial or boundary conditions, and the governing PDE through automatic differentiation. The specific equations, domains, and loss weights are given with the corresponding benchmarks in Section~\ref{sec:nr}.

Separable PINNs assign a coordinate subnetwork $\phi_j(x_j)\in\mathbb R^r$ to each physical coordinate and form the rank-$r$ approximation (Ref.~\cite{cho2023separable,oh2025separable})
\begin{equation}
u_\theta(\mathbf{x})=\sum_{s=1}^{r}\prod_{j=1}^{K}\phi_{j,s}(x_j).
\label{eq:spinn_representation}
\end{equation}
On tensor-product point sets, each coordinate subnetwork is evaluated only at the distinct values of its corresponding coordinate, after which the multidimensional prediction is assembled algebraically. This reduction in repeated network evaluation does not change the cost of an individual hidden transformation: a width-$d$ dense square map still requires $\mathcal O(d^2)$ arithmetic per hidden vector. QO-SPINN targets these square hidden maps with the pyramidal orthogonal transformation described next. The forward instantiation retains the separable output in \eqref{eq:spinn_representation}, whereas the UQ instantiation uses coordinate-wise feature extraction followed by a global random-feature head.

\paragraph{\textbf{Pyramidal Orthogonal Transformations}}
\label{sec:pyramidal_background}

For a normalized vector $q\in\mathbb R^d$, unary amplitude encoding gives
$|q\rangle=\sum_{i=1}^{d}q_i|e_i\rangle,
$
where $|e_i\rangle$ is the single-excitation basis state at mode $i$. An RBS gate acting on neighboring unary modes implements the planar rotation
\begin{equation}
G(\theta)=\begin{bmatrix}\cos\theta&\sin\theta\\-\sin\theta&\cos\theta\end{bmatrix}.
\label{eq:rbs_rotation}
\end{equation}
A pyramidal composition of $P_d=\frac{d(d-1)}{2}$ such rotations parameterizes $W\in SO(d)$ and realizes $U_W|q\rangle=|Wq\rangle$ as described in Ref.~\cite{kerenidis2021classical,landman2022quantum}.

The pyramid contains $\Theta(d^2)$ rotations but only $2d-3$ sequential rotation timesteps because gates on disjoint neighboring modes execute in parallel. Its circuit depth is therefore
\begin{equation}
D_{\rm pyr}(d)=\Theta(d),
\label{eq:pyramid_depth}
\end{equation}
and linear-depth unary state preparation does not change this asymptotic order. The same rotation sequence can be evaluated classically with
\begin{equation}
T_{\rm pyr}^{\rm classical}(d)=\mathcal O(d^2),
\label{eq:classical_pyramid_cost}
\end{equation}
for both forward evaluation and backpropagation through all $\Theta(d^2)$ rotation parameters (Ref.~\cite{landman2022quantum}). Thus the pyramidal parametrization provides exact orthogonality with quadratic classical complexity, while its quantum realization reduces the sequential circuit depth of the learned transformation to $\Theta(d)$ before readout.

\paragraph{\textbf{Signed Readout and Quantum Inference Cost}}
\label{sec:readout_model}

Because nonlinearities between QO maps are evaluated classically, each quantum output must be reconstructed with signed amplitudes; computational-basis probabilities alone are insufficient. We therefore adopt the signed $\ell_\infty$ tomography access model used in quantum neural-network analyses such as Ref.~\cite{kerenidis2019quantum,landman2022quantum}. For fixed normalized $q$, orthogonal $W$, tolerance $\epsilon>0$, and failure probability $\eta_{\rm call}$, we assume $
\Pr\!\left[\|\widehat q_W-Wq\|_\infty\leq\epsilon\right]\geq1-\eta_{\rm call}$ and
\begin{equation} 
N_{\rm tomo}=\mathcal O\!\left(\frac{\log(d/\eta_{\rm call})}{\epsilon^2}\right).
\label{eq:signed_tomography_assumption}
\end{equation}
One state-preparation-plus-pyramid execution has $\Theta(d)$ depth. Charging the tomography repetitions sequentially therefore gives
\begin{equation}
T_{\rm pyr}^{\rm quantum}=\mathcal O\!\left(\frac{d\log(d/\eta_{\rm call})}{\epsilon^2}\right)=\widetilde{\mathcal O}\!\left(\frac d{\epsilon^2}\right)
\label{eq:quantum_layer_cost}
\end{equation}
while conventional direct dense matrix--vector multiplication costs $T_{\rm dense}^{\rm classical}=\Theta(d^2)$.
Here $\epsilon$ is a chosen absolute componentwise readout tolerance; the tomography assumption does not require it to be independent of width. This is a sequential depth/access comparison, not a comparison of total work or measured execution time: the pyramid contains $\Theta(d^2)$ RBS gates, and tomography contributes the $\epsilon^{-2}$ factor. Section~\ref{sec:finite_readout_theory} links the sufficient tolerance to a prescribed physical-coordinate resolution and to width-dependent trained-model constants.

\paragraph{\textbf{Scope of the comparison}}
\label{sec:computational_scope}

The comparison in Eq~\eqref{eq:quantum_layer_cost} concerns execution of a learned orthogonal hidden map rather than end-to-end quantum training. For $B$ independently processed vectors, the cost scales by $B$ unless additional hardware parallelism is available, while SPINN reduces the number of coordinate-network evaluations on tensor-product domains. The experiments below use exact classical RBS simulation, so their wall-clock time characterizes the simulator rather than quantum hardware. An analysis considering noise, error correction, physical tomography constants, and the practical classical--quantum crossover is therefore left for future work.

\section{QO-SPINN Architectures}
\label{sec:architecture}

The norm-restored QO map is used in two architectures that answer different questions. For forward PDE regression, QO-FiLM combines the orthogonal transformation with norm-conditioned FiLM blocks and the standard separable SPINN output. To study representation geometry and uncertainty, we instead use a norm-controlled residual architecture with contractive coordinate subnetworks and a random-feature linear head. The two models are trained independently: the execution model applies to every QO hidden transformation, while the geometric guarantees rely specifically on the residual construction.

\paragraph{\textbf{Coordinate Encoding}}
\label{sec:coordinate_encoding}

For the norm-controlled architecture, we nondimensionalize each physical coordinate as
\begin{equation}
\bar{x}_j=\frac{x_j-x_{j,\min}}{x_{j,\max}-x_{j,\min}}\in[0,1].
\label{eq:nondimensional_coordinate}
\end{equation}
Throughout the geometric and finite-readout analysis, coordinate separations and resolution scales such as $\Delta$ are measured in Euclidean distance on the nondimensionalized coordinate vector $\bar{\mathbf{x}}$, not in the original dimensional coordinate units. We augment $\bar{x}_j$ with fixed Fourier features
\begin{equation}
\psi_j(\bar{x})=\frac{1}{\sqrt{p_j}}\begin{bmatrix}\cos(\xi_{j,1}\bar{x}+\nu_{j,1})\\ \sin(\xi_{j,1}\bar{x}+\nu_{j,1})\\ \vdots\\ \cos(\xi_{j,p_j}\bar{x}+\nu_{j,p_j})\\ \sin(\xi_{j,p_j}\bar{x}+\nu_{j,p_j})\end{bmatrix},
\label{eq:input_fourier_features}
\end{equation}
whose frequencies and phases are fixed after initialization and satisfy $\|\psi_j(\bar{x})\|_2=1$. The resulting width-$d$ embedding is
$E_j(\bar{x}_j)=\left[\kappa_j\bar{x}_j\;\middle\|\;\psi_j(\bar{x}_j)\;\middle\|\;0\right]\in\mathbb R^d$, where $\kappa_j>0$ with zero padding when $d>2p_j+1$. The explicit coordinate component is essential for the later lower distance bound: the periodic Fourier features enrich the representation but cannot by themselves guarantee global separation of distinct physical coordinates.

\paragraph{\textbf{Norm-Restored Quantum Orthogonal Layer}}
\label{sec:norm_restored_layer}

Let $h\in\mathbb R^d$ be a hidden state and define its magnitude and direction by
\begin{equation}
n(h)=\|h\|_2,\qquad q(h)=\begin{cases}h/\|h\|_2,&h\neq0,\\0,&h=0.\end{cases}
\label{eq:hidden_norm_direction}
\end{equation}
For $h\neq0$, $q(h)$ is unary encoded and transformed by the pyramidal circuit representing $W\in SO(d)$. Retaining $n(h)$ classically, finite signed readout defines
\begin{equation}
\widehat{\mathcal R}_W(h)=\begin{cases}n(h)\widehat q_W(h),&h\neq0,\\0,&h=0.\end{cases}
\label{eq:finite_restored_map}
\end{equation}
Under exact readout,
\begin{equation}
\mathcal R_W(h)=n(h)Wq(h)=Wh,\qquad \forall h\in\mathbb R^d.
\label{eq:exact_restoration_identity}
\end{equation}
Hence unary normalization does not alter the represented hidden transformation, and the exact restored map remains linear and differentiable at $h=0$. If the directional reconstruction satisfies $\|\widehat q_W(h)-Wq(h)\|_\infty\leq\epsilon$, then
\begin{equation}
\|\widehat{\mathcal R}_W(h)-Wh\|_2\leq\|h\|_2\sqrt d\,\epsilon.
\label{eq:restored_readout_error}
\end{equation}
The norm computation and classical preprocessing each require $\mathcal O(d)$ work and therefore do not alter the fixed-precision width scaling of the quantum execution model.

\paragraph{\textbf{Forward Regression with Norm-Conditioned FiLM}}
\label{sec:forward_film}

For forward PDE regression, we embed the restored QO map in a norm-conditioned FiLM block. Let $s(h)=~\|h\|_2^2$ and $a(s)=s/(1+s)$, and let $\gamma_{\rm F}(s),\beta_{\rm F}(s)\in\mathbb R^d$ be bounded smooth functions. We define
\begin{equation}
F_W(h)=\left(1+\gamma_{\rm F}(s(h))\right)\odot\mathcal R_W(h)+a(s(h))\beta_{\rm F}(s(h))+b.
\label{eq:forward_film}
\end{equation}
Using \eqref{eq:exact_restoration_identity}, exact readout gives $F_W(h)=~\left(1+\gamma_{\rm F}(s(h))\right)~\odot~Wh+a(s(h))\beta_{\rm F}(s(h))+b$. The dependence on $\|h\|_2^2$ is smooth at the origin, and the resulting coordinate subnetworks produce the rank-$r$ factors in \eqref{eq:spinn_representation}. This architecture is used for the forward-regression experiments The global representation guarantees developed later require the residual construction introduced next.

\paragraph{\textbf{Norm-Controlled Residual Model}}
\label{sec:qo_residual_network}

For the norm-controlled architecture, each coordinate is processed by an independent residual subnetwork with $h_{j,0}=E_j(\bar{x}_j)$ and
\begin{equation}
h_{j,\ell}=T_{j,\ell}(h_{j,\ell-1})=h_{j,\ell-1}+\rho_{j,\ell}\sigma\!\left(\mathcal R_{W_{j,\ell}}(h_{j,\ell-1})+b_{j,\ell}\right).
\label{eq:qo_residual_block}
\end{equation}
We assume that $\sigma$ is $L_\sigma$-Lipschitz and parameterize $\rho_{j,\ell}=~\rho_{\max}\tanh(\theta_{j,\ell}) $ and $ 0<\rho_{\max}<1/L_\sigma.$
Because $\|W_{j,\ell}\|_2=1$, the residual branch has Lipschitz constant
\begin{equation}
\alpha_{j,\ell}:=|\rho_{j,\ell}|L_\sigma<1.
\label{eq:qo_alpha}
\end{equation}
The contraction condition applies to the residual branch, not to the complete map $T_{j,\ell}$. In the experiments we use $\sigma(z)=~\sin z$, for which $L_\sigma=1$ and $\alpha_{j,\ell}=|\rho_{j,\ell}|$.

After $L_j$ blocks, define
\begin{equation}
z_j(\bar{x}_j)=\Phi_j(E_j(\bar{x}_j)),\qquad \Phi_j=T_{j,L_j}\circ\cdots\circ T_{j,1},
\label{eq:coordinate_representation}
\end{equation}
and concatenate the coordinate representations as
\begin{equation}
Z(\bar{\mathbf{x}})=\left[z_1(\bar{x}_1)\;\middle\|\;\cdots\;\middle\|\;z_K(\bar{x}_K)\right]\in\mathbb R^{Kd}.
\label{eq:complete_latent_representation}
\end{equation}
The representation $Z$ contains no learned global mixing; the random-feature projection used for prediction is applied only afterward.

\paragraph{\textbf{Random-Feature Prediction and Uncertainty Head}}
\label{sec:separable_rff}

The norm-controlled architecture maps the concatenated representation $Z\in\mathbb R^{Kd}$ to fixed random Fourier features (RFFs) (Ref.~\cite{rahimi2007randomfeatures}),
\begin{equation}
\varphi(Z)=\sqrt{\frac{2}{D_L}}\cos\!\left(\sqrt{2\gamma}\,\Omega_LZ+b_L\right),
\label{eq:joint_feature_vector}
\end{equation}
where the rows of $\Omega_L\in\mathbb R^{D_L\times Kd}$ are sampled independently from $\mathcal N(0,I_{Kd})$ and the entries of $b_L$ from $\mathrm{Unif}[0,2\pi]$, then fixed. Writing $\varphi(\mathbf{x})=\varphi(Z(\bar{\mathbf{x}}))$, the deterministic prediction is
\begin{equation}
\mu_\theta(\mathbf{x})=\varphi(\mathbf{x})^T\beta.
\label{eq:qo_spinn_mean}
\end{equation}
These features approximate the latent RBF kernel
$
k_Z(\mathbf{x},\mathbf{x}')=~\exp\!\left(-\gamma\|Z(\bar{\mathbf{x}})-Z(\bar{\mathbf{x}}')\|_2^2\right)$.
The geometric results below concern the deterministic representation $Z$ and this exact kernel; no global bi-Lipschitz property is assumed for a finite random-feature realization.

\paragraph{\textbf{Physics-Informed Training}}
\label{sec:unified_training}

Both architectures are trained with the physics-informed objective \eqref{eq:generic_pinn_loss},
\begin{equation}
\widehat\Theta=\arg\min_{\Theta}\mathcal L(\Theta),
\label{eq:qo_training_objective}
\end{equation}
using their respective deterministic outputs: the separable prediction \eqref{eq:spinn_representation} for QO-FiLM and the random-feature mean \eqref{eq:qo_spinn_mean} for the norm-controlled model. All PDE derivatives are taken through these deterministic predictions.

\paragraph{\textbf{Uncertainty from the Same Trained Model}}
\label{sec:unified_uq}

Distance-aware single-model UQ through spectrally controlled representations and GP/RFF heads is closely related to SNGP as defined in Ref.~\cite{liu2020sngp,liu2023simple}. In the PINN setting, Li et al. Recent work (Ref.~\cite{li2024pinnuq}) combine spectral normalization of hidden layers with a Gaussian-process output layer to obtain distance-aware predictive uncertainty. Our construction follows the same broad distance-awareness principle but differs in that unit operator norm is structural for QO layers, the deterministic physics-informed predictor is trained first and then frozen, and uncertainty is subsequently introduced through a post-hoc Bayesian-linear covariance. A complementary recent approach embeds PDE information directly into the prior of a Bayesian neural network (Ref.~\cite{menicali2025pibnn}); in contrast, the covariance used here is conditional on a deterministic physics-informed representation and is not the posterior of a fully Bayesian PINN. Recent work has also developed analytical propagation of aleatoric measurement uncertainty through trained neural networks (Ref.~\cite{jungmann2024analytical}); this addresses a different uncertainty source from the representation-dependent epistemic covariance considered here. Bayesian last-layer models provide an analytically tractable basis for this type of post-hoc construction (Ref.~\cite{fiedler2023bll}). Here, after physics-informed training, the norm-controlled representation, random-feature realization, and coefficient vector $\widehat\beta$ are fixed. Let $\Phi_y\in\mathbb R^{N_y\times D_L}$ denote the supervised design matrix whose $i$th row is $\varphi(\mathbf{x}^{(i)})^T$, so that $\Phi_y$ collects the features at supervised observation locations. PDE collocation points affect the trained mean through \eqref{eq:qo_training_objective} but are not treated as observed targets in the covariance. For prior precision $\lambda>0$ and observation variance $\sigma_y^2>0$, define
\begin{equation}
\Sigma_\beta=\left(\lambda I+\sigma_y^{-2}\Phi_y^T\Phi_y\right)^{-1}.
\label{eq:posthoc_covariance_unified}
\end{equation}
For $\varphi_*=\varphi(\mathbf{x}_*)$,
\begin{equation}
\sigma_{\rm epi,*}^2=\varphi_*^T\Sigma_\beta\varphi_*,\qquad \sigma_{\rm pred,*}^2=\sigma_{\rm epi,*}^2+\sigma_y^2.
\label{eq:predictive_variance_unified}
\end{equation}
The predictive mean remains $\mu_*=\varphi_*^T\widehat\beta$; uncertainty therefore augments the trained predictor without retraining the feature extractor or replacing its coefficients. Since $\widehat\beta$ is learned from the full physics-informed objective whereas $\Sigma_\beta$ uses supervised observations, this construction is a post-hoc Bayesian-linear covariance conditional on the learned representation rather than the posterior of a fully Bayesian PINN. When $N_y\ll D_L$, the Woodbury identity gives
\begin{equation}
\Sigma_\beta=\lambda^{-1}I-\lambda^{-2}\Phi_y^T\left(\sigma_y^2I+\lambda^{-1}\Phi_y\Phi_y^T\right)^{-1}\Phi_y,
\label{eq:woodbury_covariance}
\end{equation}
so the nontrivial factorization is performed in the $N_y$-dimensional observation space.

\paragraph{\textbf{Matched Dense Controls}}
\label{sec:matched_dense_architecture}

Each architecture is compared with a matched dense control that preserves all non-map components and replaces only the square QO transformation. In the forward study, QO-FiLM is matched with an unconstrained dense-FiLM model using the same coordinate features, width, depth, separable output, loss, optimization protocol, and paired initialization. In the norm-controlled study, the dense model uses the same embedding, residual architecture, random-feature realization, training data, and covariance construction, but replaces each QO map by
\begin{equation}
W_{j,\ell}^{\rm dense}=\frac{A_{j,\ell}}{\sigma_{\max}(A_{j,\ell})},\qquad A_{j,\ell}\neq0.
\label{eq:matched_dense_normalization}
\end{equation}
Exact spectral normalization gives $\|W_{j,\ell}^{\rm dense}\|_2=1$ and therefore the same sufficient residual-branch norm condition as QO. The distinction is enforcement: QO satisfies unit operator norm structurally, whereas the dense model must compute or estimate the normalization scale.
The matched controls therefore differ only in how the square hidden maps are parameterized and constrained. Since the learned RBS maps are evaluated by exact classical simulation in all predictive experiments, differences between QO and dense models reflect parametrization and optimization rather than quantum-hardware effects. The quantum comparison concerns the alternative execution model for the learned orthogonal map.

\section{Geometry, Readout, and Computational Scaling}
\label{sec:theory}

The QO construction links computational structure and representation geometry through the same unit-norm hidden map. We first extend the single-layer execution cost to repeated hidden-map evaluations. In the norm-controlled residual model, exact orthogonality then gives layer-wise two-sided bounds that compose into a global bi-Lipschitz representation bound and corresponding latent-kernel bounds. We subsequently examine how approximate dense spectral normalization and finite quantum readout alter these guarantees before considering the scale of the post-hoc covariance.

\paragraph{\textbf{Repeated Hidden-Map Complexity}}
\label{sec:hidden_map_scaling}

Let $H$ denote the number of width-$d$ square hidden-map evaluations in a deterministic workload. Direct dense multiplication and exact classical pyramidal execution require
\begin{equation}
T_{\rm hidden}^{\rm dense}=\Theta(Hd^2),\qquad T_{\rm hidden}^{\rm QO,classical}=\mathcal O(Hd^2).
\label{eq:repeated_classical_cost}
\end{equation}
If each quantum readout is reconstructed to componentwise tolerance $\epsilon$ and all $H$ calls are required to succeed with probability at least $1-\eta$, assigning failure probability $\eta/H$ per call gives
\begin{equation}
T_{\rm hidden}^{\rm QO,quantum}=\mathcal O\!\left(\frac{Hd\log(dH/\eta)}{\epsilon^2}\right)=\widetilde{\mathcal O}\!\left(\frac{Hd}{\epsilon^2}\right).
\label{eq:repeated_quantum_cost}
\end{equation}
For a width-indexed family of workloads, let $H=H(d)$ and allow the requested componentwise precision $\epsilon=\epsilon(d)$ to vary with width. Relative to conventional direct dense multiplication, the quantum access-cost upper bound in \eqref{eq:repeated_quantum_cost} is $o(H(d)d^2)$ whenever $\epsilon(d)^{-2}\log(dH(d)/\eta)=o(d)$. For fixed $\epsilon>0$, fixed $\eta$, and $H(d)$ growing at most polynomially with $d$, the ratio of this upper bound to the conventional dense cost is $\mathcal O(\log d/d)$, giving a nearly linear nominal separation up to logarithmic factors. This is a conditional comparison of access-cost bounds; no finite-width crossover or hardware acceleration is demonstrated. This comparison isolates the shared square hidden transformations; architecture-specific output computations are additional. Classically, the pyramidal parametrization also preserves exact orthogonality with $\mathcal O(d^2)$ forward and backward work, whereas recomputing an exact dense SVD normalization requires cubic matrix factorization.

\paragraph{\textbf{Exact-Readout Representation Geometry}} \label{sec:exact_geometry} Analytical Lipschitz bounds provide a general tool for characterizing sensitivity of neural representations; recent work, for example, derives analytical upper bounds on local Lipschitz constants of feedforward ReLU networks as in Ref.~\cite{avant2023analytical}. The structure considered here permits a different guarantee: global two-sided bounds on the learned coordinate representation, including a strictly positive lower bound under the stated residual-branch condition. For the embedding \eqref{eq:input_fourier_features}, define
\begin{equation}
L_{\psi_j}:=\sup_{\bar x\in[0,1]}\|\psi_j'(\bar x)\|_2=\left(\frac{1}{p_j}\sum_{s=1}^{p_j}\xi_{j,s}^2\right)^{1/2}.
\label{eq:fourier_lipschitz_constant}
\end{equation}

\begin{lemma}[Coordinate embedding]
\label{lem:coordinate_embedding_new}
For all $\bar x,\bar y\in[0,1]$,
\begin{equation}
\kappa_j|\bar x-\bar y|\leq\|E_j(\bar x)-E_j(\bar y)\|_2\leq\sqrt{\kappa_j^2+L_{\psi_j}^2}\,|\bar x-\bar y|.
\label{eq:coordinate_embedding_bounds_new}
\end{equation}
\end{lemma}

For a residual block $T_{j,\ell}(h)=h+\rho_{j,\ell}\sigma(W_{j,\ell}h+b_{j,\ell})$, define
$\alpha_{j,\ell}:=|\rho_{j,\ell}|L_\sigma\|W_{j,\ell}\|_2.$

\begin{lemma}[Two-sided residual-block bound]
\label{lem:residual_block_new}
If $\alpha_{j,\ell}<1$, then for all $h,h'\in\mathbb R^d$,
\begin{equation}
(1-\alpha_{j,\ell})\|h-h'\|_2
\!\leq\!
\|T_{j,\ell}(h)-T_{j,\ell}(h')\|_2
\!\leq\!
(1+\alpha_{j,\ell})\|h-h'\|_2.
\label{eq:residual_block_bounds_new}
\end{equation}
Moreover, $T_{j,\ell}$ is bijective and $T_{j,\ell}^{-1}$ is $(1-\alpha_{j,\ell})^{-1}$-Lipschitz.
\end{lemma}

For exact QO layers, $\|W_{j,\ell}\|_2=1$ and therefore $\alpha_{j,\ell}=|\rho_{j,\ell}|L_\sigma<1$. Define
$m_{\Phi_j}:=\prod_{\ell=1}^{L_j}(1-\alpha_{j,\ell}) $ and $ M_{\Phi_j}:=~\prod_{\ell=1}^{L_j}(1+\alpha_{j,\ell}),
$
and $m_j:=\kappa_jm_{\Phi_j},$ and $ M_j:=~\sqrt{\kappa_j^2+L_{\psi_j}^2}\,M_{\Phi_j}.$

\begin{theorem}[Exact-readout representation geometry]
\label{thm:complete_geometry_new}
If every residual branch satisfies $\alpha_{j,\ell}<1$, then
\begin{equation}
m_j|\bar x_j-\bar y_j|\leq\|z_j(\bar x_j)-z_j(\bar y_j)\|_2\leq M_j|\bar x_j-\bar y_j|.
\label{eq:coordinate_representation_bounds_new}
\end{equation}
For the concatenated representation $Z$,
\begin{equation}
m_Z\|\bar{\mathbf{x}}-\bar{\mathbf{y}}\|_2\leq\|Z(\bar{\mathbf{x}})-Z(\bar{\mathbf{y}})\|_2\leq M_Z\|\bar{\mathbf{x}}-\bar{\mathbf{y}}\|_2,
\label{eq:complete_geometry_bounds_new}
\end{equation}
where
$m_Z:=\min_jm_j>0,\qquad M_Z:=\max_jM_j, $ and $ \chi_Z:=\frac{M_Z}{m_Z}.
$
Hence $Z$ is bi-Lipschitz onto its image.
\end{theorem}

The ratio $\chi_Z=M_Z/m_Z$ summarizes the worst-case distortion allowed by these bounds: smaller values indicate tighter sufficient control of latent distances. This geometric quantity does not by itself determine predictive error or probabilistic calibration, which depend on additional properties of the learned predictor.

\paragraph{\textbf{Structural QO Norm Control and Dense Spectral Normalization}}
\label{sec:qovs_dense_certificate}

The exact dense normalization in \eqref{eq:matched_dense_normalization} also gives $\|W_{j,\ell}^{\rm dense}\|_2=1$ and therefore satisfies the same sufficient residual-block condition as exact QO for equal residual scales. The difference is how unit operator norm is enforced. For an estimated scale $\widehat s_{j,\ell}>0$, define
$\widetilde W_{j,\ell}:=\frac{A_{j,\ell}}{\widehat s_{j,\ell}},$ and $ \tau_{j,\ell}:=\frac{\sigma_{\max}(A_{j,\ell})}{\widehat s_{j,\ell}}=\|\widetilde W_{j,\ell}\|_2.$

\begin{corollary}[Certificate under inexact spectral normalization]
\label{cor:inexact_sn_new}
If $\tau_{j,\ell}\leq1+\xi_{j,\ell}$, then the residual branch has Lipschitz constant at most
\begin{equation}
\bar\alpha_{j,\ell}:=|\rho_{j,\ell}|L_\sigma(1+\xi_{j,\ell}).
\end{equation}
If $\bar\alpha_{j,\ell}<1$ for every block, Theorem~\ref{thm:complete_geometry_new} holds with $\alpha_{j,\ell}$ replaced by $\bar\alpha_{j,\ell}$ in the sufficient constants.
\end{corollary}

Thus underestimating the spectral norm, for which $\tau_{j,\ell}>1$, directly loosens the representation certificate, whereas exact QO fixes $\tau_{j,\ell}=1$ by construction.

\paragraph{\textbf{Exact Latent-Kernel Consequences}}
\label{sec:kernel_theory}

For the exact latent kernel $k_Z(\mathbf{x},\mathbf{y})=\exp[-\gamma\|Z(\bar{\mathbf{x}})-Z(\bar{\mathbf{y}})\|_2^2]$, Theorem~\ref{thm:complete_geometry_new} gives the following consequence.

\begin{corollary}[Latent-kernel geometry]
\label{cor:kernel_geometry_new}
Under the assumptions of Theorem~\ref{thm:complete_geometry_new},
\begin{equation}
\exp\!\left[-\gamma M_Z^2\|\bar{\mathbf{x}}-\bar{\mathbf{y}}\|_2^2\right]\leq k_Z(\mathbf{x},\mathbf{y})\leq\exp\!\left[-\gamma m_Z^2\|\bar{\mathbf{x}}-\bar{\mathbf{y}}\|_2^2\right].
\label{eq:kernel_geometry_bounds_new}
\end{equation}
Because $Z$ is a coordinate-wise concatenation, $k_Z(\mathbf{x},\mathbf{y})=\prod_{j=1}^{K}\exp[-\gamma\|z_j(\bar x_j)-z_j(\bar y_j)\|_2^2]$.
\end{corollary}

The corollary concerns the exact kernel induced by $Z$; the finite random-feature approximation is not assumed to be globally bi-Lipschitz.

\paragraph{\textbf{Finite-Readout Geometry}}
\label{sec:finite_readout_theory}

The local error in \eqref{eq:restored_readout_error} accumulates as the perturbed state passes through successive residual blocks. To bound this accumulation, assume $|\sigma(a)_i|\leq S_\sigma$; for the sine activation, $S_\sigma=1$. Since $\|\psi_j(\bar x_j)\|_2=1$, define
$
B_{j,0}:=~\sqrt{\kappa_j^2+1} $ and $ B_{j,\ell}:=~B_{j,0}+S_\sigma\sqrt d\sum_{r=1}^{\ell}|\rho_{j,r}|.
$
For componentwise readout tolerance $\epsilon_{j,\ell}$, let
$
\delta_{j,\ell}:=~\rho_{j,\ell}|L_\sigma B_{j,\ell-1}\sqrt d\,\epsilon_{j,\ell}, $ and $ D_j:=~\sum_{\ell=1}^{L_j}\delta_{j,\ell}\prod_{r=\ell+1}^{L_j}(1+\alpha_{j,r}),$
and
$
D_Z:=~\left(\sum_{j=1}^{K}D_j^2\right)^{1/2}$.

\begin{theorem}[Finite-readout distance distortion]
\label{thm:finite_readout_new}
Let $Q:=~\sum_{j=1}^{K}L_j$ denote the number of QO readout calls required for one complete UQ feature-extractor evaluation. For a fixed input pair $(\bar{\mathbf{x}},\bar{\mathbf{y}})$, suppose each of the corresponding $2Q$ calls satisfies its componentwise readout event with conditional failure probability at most $\eta/(2Q)$. Then, with probability at least $1-\eta$,
\begin{equation}
m_Z\|\bar{\mathbf{x}}-\bar{\mathbf{y}}\|_2-2D_Z\leq\|\widehat Z(\bar{\mathbf{x}})-\widehat Z(\bar{\mathbf{y}})\|_2\leq M_Z\|\bar{\mathbf{x}}-\bar{\mathbf{y}}\|_2+2D_Z.
\label{eq:finite_readout_geometry_new}
\end{equation}
\end{theorem}

For a common componentwise tolerance $\epsilon_{j,\ell}=\epsilon$, write
$D_Z=~C_Z\epsilon,$
with
$C_j:=\sum_{\ell=1}^{L_j}|\rho_{j,\ell}|L_\sigma B_{j,\ell-1}\sqrt d\prod_{r=\ell+1}^{L_j}(1+\alpha_{j,r})$
and $C_Z:=\left(\sum_{j=1}^{K}C_j^2\right)^{1/2}$.
This gives the sufficient readout-dependent resolution
$
r_{\rm read}:=\frac{2C_Z\epsilon}{m_Z}.$

\begin{corollary}[Readout precision for a prescribed physical-coordinate scale]
\label{cor:resolution_precision_new}
Fix $\Delta>0$ and $\zeta\in(0,1)$. If
\begin{equation}
\epsilon\leq\frac{(1-\zeta)m_Z\Delta}{2C_Z},
\label{eq:epsilon_for_resolution_new}
\end{equation}
then every fixed queried pair satisfying $\|\bar{\mathbf{x}}-\bar{\mathbf{y}}\|_2\geq\Delta$ obeys
\begin{equation}
\|\widehat Z(\bar{\mathbf{x}})-\widehat Z(\bar{\mathbf{y}})\|_2\geq\zeta m_Z\Delta
\end{equation}
on its joint readout-success event. Under the signed-readout model in \eqref{eq:signed_tomography_assumption}, a sufficient per-readout repetition budget therefore scales as
\begin{equation}
N_{\rm tomo}=\mathcal{O}\!\left(\frac{C_Z^2}{(1-\zeta)^2m_Z^2\Delta^2}\log\!\frac{Qd}{\eta}\right),
\label{eq:resolution_call_scaling_new}
\end{equation}
up to access-model constants.
\end{corollary}

The resulting tomography budget scales as $\Delta^{-2}$: certifying finer physical-coordinate resolution requires proportionally smaller componentwise readout error. Combining \eqref{eq:repeated_quantum_cost} with \eqref{eq:epsilon_for_resolution_new}, a sufficient access cost for $H$ hidden-map calls at prescribed resolution $\Delta$ is
\begin{equation}
T_{\rm hidden,\Delta}^{\rm QO,quantum}=\mathcal O\!\left(\frac{Hd\,C_Z^2}{(1-\zeta)^2m_Z^2\Delta^2}\log\!\frac{dH}{\eta}\right),
\label{eq:fixed_resolution_quantum_cost}
\end{equation}
up to access-model constants. For a fixed resolution $\Delta>0$, retention factor $\zeta\in(0,1)$, and failure target $\eta\in(0,1)$, the sufficient tolerance is $\epsilon_{\max}(d;\Delta,\zeta):=(1-\zeta)m_Z(d)\Delta/[2C_Z(d)]$ when $C_Z(d)>0$. Consequently, the bound in \eqref{eq:fixed_resolution_quantum_cost} is $o(H(d)d^2)$ if
\begin{equation}
\frac{C_Z(d)^2\log(dH(d)/\eta)}{m_Z(d)^2d\Delta^2}\longrightarrow0.
\label{eq:fixed_resolution_access_criterion}
\end{equation}
The displayed quantity is computable for each trained residual model, but its asymptotic behavior requires studying a family of models across increasing widths. Moreover, letting $\rho_*(d):=\max_{j,\ell}|\rho_{j,\ell}(d)|$, the definitions of $C_Z$, $m_Z$, and $B_{j,\ell-1}$ imply $C_Z(d)/m_Z(d)\geq\sqrt d\,L_\sigma\rho_*(d)$. Therefore, if $\rho_*(d)$ remains bounded away from zero, this sufficient fixed-resolution condition fails. This limitation pertains to the present worst-case certificate and does not rule out sharper task-specific bounds or improved quantum implementations.

\paragraph{\textbf{Marginal Cost and Scale of the Post-Hoc Covariance}}
\label{sec:uq_theory}

For $N_y\ll D_L$, let $A_y:=\sigma_y^2I_{N_y}+\lambda^{-1}\Phi_y\Phi_y^T$. Using the Woodbury form in \eqref{eq:woodbury_covariance}, the post-hoc covariance requires
$
T_{\rm UQ,setup}=\mathcal O(N_y^2D_L+N_y^3) $ and $ T_{\rm UQ,query}=\mathcal O(N_yD_L+N_y^2),
$
after deterministic features have been computed. Thus uncertainty reuses the trained representation without an additional feature extractor.

Let $G:=\Phi_y^T\Phi_y\succeq0$, $\tau:=\sigma_y^2$, and $\Sigma(\lambda,\tau):=(\lambda I+\tau^{-1}G)^{-1}$.

\begin{theorem}[Post-hoc covariance scale]
\label{thm:covariance_scale_new}
For $\lambda,\tau>0$,
\begin{equation}
0\prec\Sigma(\lambda,\tau)\preceq\lambda^{-1}I,\qquad 0\leq\sigma_{\rm epi,*}^2\leq\frac{\|\varphi_*\|_2^2}{\lambda}.
\label{eq:covariance_order_new}
\end{equation}
Moreover,
\begin{equation}
\frac{\partial\Sigma}{\partial\lambda}=-\Sigma^2\preceq0,\qquad \frac{\partial\Sigma}{\partial\tau}=\tau^{-2}\Sigma G\Sigma\succeq0.
\label{eq:covariance_derivatives_new}
\end{equation}
Thus epistemic variance decreases with $\lambda$ and increases with $\tau$, while predictive variance $\sigma_{\rm pred,*}^2=\sigma_{\rm epi,*}^2+\tau$ increases strictly with $\tau$.
\end{theorem}
Since \eqref{eq:joint_feature_vector} gives $\|\varphi(\mathbf{x})\|_2^2\leq2$, $
\sigma_{\rm epi}^2(\mathbf{x})\leq\frac{2}{\lambda}$ and $ \sigma_{\rm pred}^2(\mathbf{x})\leq\frac{2}{\lambda}+\sigma_y^2$.

\begin{corollary}[Representation geometry does not determine calibration]
\label{cor:geometry_not_calibration_new}
The constants $m_Z$, $M_Z$, and $\chi_Z$ and the exact latent-kernel bounds do not imply a positive lower bound on epistemic variance or nominal predictive coverage. In particular, with $Z$, $\Phi_y$, and $\varphi_*$ fixed then $\lambda\rightarrow\infty$ implies $\Sigma_\beta\rightarrow0$ and $\sigma_{\rm epi,*}^2\rightarrow0$.
while the deterministic representation geometry remains unchanged.
\end{corollary}

The resulting picture separates three effects. The pyramidal execution bound belongs to the QO transformation itself; the bi-Lipschitz, kernel, and finite-readout bounds arise from placing that transformation inside the norm-controlled residual architecture; and the post-hoc covariance introduces an uncertainty scale through $(\lambda,\sigma_y^2)$ without changing the deterministic representation. The experiments below examine these effects separately.

\section{Numerical Experiments}
\label{sec:nr}

The experiments address three questions: 1) whether QO hidden maps remain competitive for forward PDE regression, 2) whether the geometric and finite-readout bounds are reflected in the trained representations, and 3) how uncertainty behaves when the norm-controlled model is transferred from Heat to Burgers. Unless stated otherwise, each metric or derived quantity is computed separately for each seed before reporting the arithmetic mean and population standard deviation; nonlinear quantities such as $\chi_Z$, CE, $r_{\rm read}/\epsilon$, and $\epsilon_{\max}$ are therefore not reconstructed from averaged components. All predictive experiments evaluate QO transformations by exact classical RBS simulation, while bounded readout perturbations are introduced only in the finite-readout study. Proofs, hyperparameters, seed-level results, ablations, detailed resource accounting, contextual forward and UQ baselines, and extended sensitivity results are given in the supplementary material.

\begin{table*}[t]
\centering
\caption{Architecture and square-map resource proxies for the decisive controls. QO and dense models are matched in width, depth, data, non-map components, and optimization, but not in square-map parameter count because $SO(d)$ has $d(d-1)/2$ rotational degrees of freedom whereas an unconstrained dense $d\times d$ map has $d^2$ entries. Quantum quantities are per QO map and include one tomography ancilla. External PINN, SPINN, XPINN, and PIKAN results reported in the supplementary material are contextual rather than matched controls.}
\label{tab:matched_resource_setup}
\setlength{\tabcolsep}{4pt}
\resizebox{\textwidth}{!}{%
\begin{tabular}{l l c c c c c c}
\hline
Study & Model & Width & Square maps/coord. & Output/head & Square-map parameters & Qubits/map & RBS map/load \\
\hline
Forward & Dir.-only RBS & 20 & 2 & CP $r=32/16$ & 190 angles & 21 & $190/19$ \\
Forward & Dense-FiLM & 20 & 2 & CP $r=32/16$ & 400 weights & -- & -- \\
Forward & QO-FiLM & 20 & 2 & CP $r=32/16$ & 190 angles & 21 & $190/19$ \\
UQ & Dense-SN & 35 & 3 & $D_L=1024$ RFF & 1225 weights & -- & -- \\
UQ & QO & 35 & 3 & $D_L=1024$ RFF & 595 angles & 36 & $595/34$ \\
\hline
\end{tabular}}
\end{table*}

\paragraph{\textbf{Forward PDE Benchmarks}}
\label{sec:forward_benchmarks}

We compare QO-FiLM with a matched dense-FiLM model on viscous Burgers and one-, two-, and three-dimensional advection--diffusion problems. Within each seed pair, the models share the data, width, depth, coordinate features, non-map initialization, optimization protocol, and initial numerical square map, leaving the hidden-map parametrization as the intended architectural difference. Table~\ref{tab:matched_forward_results} reports the matched architectural comparison, including the direction-only RBS ablation. PINN, SPINN, XPINN and PIKAN results (Ref.~\cite{raissi2019physics, cho2023separable, jagtap2020extended, jagtap2020extended, patra2025physics}) are reported in the supplementary material for task-level context. The recent methods discussed in the Introduction address different PDEs, training objectives, generalization settings, or uncertainty protocols and are therefore compared methodologically rather than treated as matched numerical baselines. Full training and evaluation details are given in the supplementary material.

\begin{table*}[t]
\centering
\caption{Matched midpoint-grid forward evaluation. Entries are mean $\pm$ population standard deviation. QO-FiLM and matched dense-FiLM use equal width and paired non-map settings; direction-only RBS is included as an architectural ablation.}
\label{tab:matched_forward_results}
\small
\setlength{\tabcolsep}{4pt}
\renewcommand{\arraystretch}{1.12}
\begin{tabular*}{\textwidth}{@{\extracolsep{\fill}}llccc@{}}
\hline
\textbf{Equation} & \textbf{Metric} & \shortstack{\textbf{Dir.-only}\\\textbf{RBS}} & \shortstack{\textbf{Matched}\\\textbf{Dense-FiLM}} & \textbf{QO-FiLM} \\
\hline
\multirow{2}{*}{Burgers (1D)} & Rel. $L_2$ & $1.07\times10^{-3}\pm1.18\times10^{-4}$ & $4.31\times10^{-4}\pm4.36\times10^{-5}$ & $5.19\times10^{-4}\pm1.58\times10^{-4}$ \\
& Max Error & $3.38\times10^{-3}\pm1.76\times10^{-4}$ & $1.14\times10^{-3}\pm7.04\times10^{-5}$ & $1.27\times10^{-3}\pm4.17\times10^{-4}$ \\
\hline
\multirow{2}{*}{Adv.-Diff. (1D)} & Rel. $L_2$ & $2.49\times10^{-3}\pm6.96\times10^{-4}$ & $3.03\times10^{-3}\pm1.25\times10^{-3}$ & $4.58\times10^{-3}\pm7.49\times10^{-4}$ \\
& Max Error & $5.28\times10^{-2}\pm1.46\times10^{-2}$ & $8.32\times10^{-2}\pm2.99\times10^{-2}$ & $1.14\times10^{-1}\pm4.09\times10^{-2}$ \\
\hline
\multirow{2}{*}{Adv.-Diff. (2D)} & Rel. $L_2$ & $2.79\times10^{-3}\pm4.17\times10^{-4}$ & $4.25\times10^{-3}\pm6.36\times10^{-4}$ & $3.08\times10^{-3}\pm5.11\times10^{-4}$ \\
& Max Error & $0.285\pm0.0798$ & $0.414\pm0.0650$ & $0.334\pm0.0773$ \\
\hline
\multirow{2}{*}{Adv.-Diff. (3D)} & Rel. $L_2$ & $0.0299\pm0.0162$ & $0.0226\pm0.00170$ & $0.0224\pm0.00198$ \\
& Max Error & $6.88\pm3.87$ & $5.52\pm0.750$ & $5.54\pm0.949$ \\
\hline
\end{tabular*}
\end{table*}

No uniform accuracy advantage emerges from the QO parametrization. QO-FiLM has lower mean relative $L_2$ error than matched dense-FiLM on 2D advection--diffusion and a slightly lower mean on the 3D problem, whereas dense-FiLM performs better on Burgers and 1D advection--diffusion. The amplitude ablation also shows that norm restoration is not, by itself, an accuracy mechanism: restoration-only performs worse than direction-only RBS in the completed Burgers and 2D runs, while adding norm-conditioned FiLM improves over restoration-only in all six paired cases. The full ablation and seed-level results are reported in the supplementary material.

\paragraph{\textbf{UQ Transfer Protocol}}
\label{sec:uq_protocol}

The remaining experiments use the norm-controlled residual/RFF model. We select the covariance parameters $(\lambda,\sigma_y^2)$ on Heat and then transfer them unchanged to Burgers, without target-domain recalibration. This deliberately tests whether the learned uncertainty structure transfers across PDEs without allowing its scale to be retuned on the target problem. Complete hyperparameters are given in the supplementary material.

The matched QO and dense UQ models use width $35$, three residual blocks per coordinate, sine activation, the same normalized-coordinate/Fourier embedding, and the same global RFF head. They differ only in the square hidden maps: exact RBS orthogonal transformations for QO and exact-SVD spectral normalization in \eqref{eq:matched_dense_normalization} for dense.

\paragraph{\textbf{Covariance-Hyperparameter Transfer from Heat to Burgers}}
\label{sec:uq_transfer_design}

Within each PDE and seed, QO and dense runs share the supervised observations, collocation and validation points, reference solution, RFF realization, Fourier features, initial residual scales, and all non-map initial tensors. Model selection, optimizer, loss weights, and training schedule are also matched between architectures within each PDE, while Heat and Burgers retain their respective PDE-specific training settings. Neither architecture is tuned independently within a PDE; the full reproducibility details are provided in the supplementary material.
Heat uses the sparse-data problem $u_t=\mu u_{xx}$ with $\mu=1/\pi^2$ on $(x,t)\in[-1,1]\times[0,1]$. Transfer is evaluated on the distinct Burgers problem $u_t+u u_x=\nu u_{xx}$ with $ \nu=0.1/\pi,$
with $u(x,0)=-\sin(\pi x)$ and homogeneous Dirichlet boundary conditions. Both architectures use identical sparse-data and optimization protocols within each PDE. Burgers is evaluated on a $101\times101$ grid against the Cole--Hopf reference. Observation counts, collocation design, optimizer settings, loss weights, and reference-solution construction are reported in the supplementary material.

\paragraph{\textbf{Representation Geometry}}
\label{sec:geometry_experiments}

We first examine whether the trained representations satisfy the analytic geometry developed in Section~\ref{sec:theory}. Table~\ref{tab:geometry_checks} reports the resulting bounds and distortion ratios for Heat and Burgers.

\begin{table}[t]
\centering
\caption{Exact latent-geometry and kernel checks on Heat and Burgers. Entries are mean $\pm$ population standard deviation. ``Geom./kernel viol.'' reports sampled latent-geometry violations and exact latent-kernel violations, respectively, beyond the fixed numerical tolerances.}
\label{tab:geometry_checks}
\small
\setlength{\tabcolsep}{2.5pt}
\renewcommand{\arraystretch}{1.15}
\begin{tabular*}{\columnwidth}{@{\extracolsep{\fill}}llcccc@{}}
\hline
PDE & Model & $m_Z$ & $M_Z$ & $\chi_Z$ & \shortstack{Geom./kernel\\viol.} \\
\hline
Heat & QO & \ms{0.33458}{0.00907} & \ms{2.48945}{0.03601} & \ms{7.44897}{0.31292} & $0/0$ \\
Heat & Dense & \ms{0.30418}{0.04758} & \ms{2.63012}{0.20064} & \ms{8.94625}{1.92451} & $0/0$ \\
Burgers & QO & \ms{0.22776}{0.04253} & \ms{3.02159}{0.26071} & \ms{14.04369}{4.18607} & $0/0$ \\
Burgers & Dense & \ms{0.16336}{0.02331} & \ms{3.44503}{0.19455} & \ms{21.77105}{4.78023} & $0/0$ \\
\hline
\end{tabular*}
\end{table}

Every trained residual branch satisfies $\alpha_{j,\ell}<1$, and all sampled latent-distance and exact latent-RBF inequalities hold within the prescribed numerical tolerances. QO has a lower mean $\chi_Z$ than the matched dense model on both PDEs, with the same ordering in two of three Heat pairs and all three Burgers pairs. The result therefore indicates tighter sufficient control of latent-distance distortion for these QO checkpoints, but it does not imply lower prediction error or better calibration.

\paragraph{\textbf{Uncertainty Under Covariance-Hyperparameter Transfer from Heat to Burgers}}
\label{sec:uq_metrics_results}

For each checkpoint we report RMSE, $\mathrm{MNLL}=N^{-1}\sum_i[\frac12\log(2\pi\sigma_i^2)+(y_i-\mu_i)^2/(2\sigma_i^2)]$, $\mathrm{PICP2}=N^{-1}\sum_i\mathbf 1\{|y_i-\mu_i|\leq2\sigma_i\}$, and $\mathrm{CE}=|\mathrm{PICP2}-0.9545|$. MPIW3 is the mean width of $\mu\pm3\sigma$; EAC is the Pearson correlation between absolute error and predictive standard deviation; DAC is the Pearson correlation between predictive standard deviation and nondimensionalized distance to the nearest supervised observation. PDE collocation points are not distance anchors.

The metrics probe complementary properties of uncertainty: DAC measures sensitivity to distance from supervised data, EAC measures association with realized prediction error, and PICP2/CE measure absolute interval calibration. MPIW3 is interpreted jointly with coverage because interval width alone does not determine calibration quality.

\begin{table*}[t]
\centering
\caption{Heat-to-Burgers uncertainty transfer with covariance parameters selected on Heat and held fixed on Burgers. Entries are mean $\pm$ population standard deviation.}
\label{tab:uq_summary}
\small
\setlength{\tabcolsep}{2.5pt}
\renewcommand{\arraystretch}{1.18}
\begin{tabular*}{\textwidth}{@{\extracolsep{\fill}}llccccccc@{}}
\hline
PDE & Model & RMSE & MNLL & PICP2 & CE & MPIW3 & EAC & DAC \\
\hline
Heat & QO & \ms{2.53920\times10^{-4}}{0.06551\times10^{-4}} & \ms{-6.7163}{0.0443} & \ms{0.88459}{0.01202} & \ms{0.06991}{0.01202} & \ms{1.132219\times10^{-3}}{0.002242\times10^{-3}} & \ms{-0.2411}{0.1194} & \ms{0.87182}{0.01190} \\
Heat & Dense & \ms{1.83491\times10^{-4}}{0.08374\times10^{-4}} & \ms{-7.1660}{0.0405} & \ms{0.95906}{0.00583} & \ms{0.00598}{0.00435} & \ms{1.141001\times10^{-3}}{0.004362\times10^{-3}} & \ms{-0.1873}{0.0811} & \ms{0.90798}{0.00873} \\
Burgers & QO & \ms{0.10157}{0.00856} & \ms{1.34263\times10^{5}}{0.22713\times10^{5}} & \ms{0.00657}{0.00265} & \ms{0.94793}{0.00265} & \ms{1.158279\times10^{-3}}{0.003352\times10^{-3}} & \ms{0.4757}{0.0549} & \ms{0.85499}{0.00409} \\
Burgers & Dense & \ms{0.08634}{0.01224} & \ms{9.70636\times10^{4}}{2.75387\times10^{4}} & \ms{0.00863}{0.00131} & \ms{0.94587}{0.00131} & \ms{1.170049\times10^{-3}}{0.002867\times10^{-3}} & \ms{0.4423}{0.0714} & \ms{0.83617}{0.01613} \\
\hline
\end{tabular*}
\end{table*}

The transfer results reveal a clear separation between the structure and the scale of uncertainty. DAC remains high when moving from Heat to Burgers, decreasing from $0.8718$ to $0.8550$ for QO and from $0.9080$ to $0.8362$ for dense, so predictive uncertainty remains strongly associated with distance from supervised observations. Absolute calibration, however, collapses: PICP2 falls from $0.8846$ and $0.9591$ on Heat to $0.00657$ and $0.00863$ on Burgers for QO and dense, respectively. At the same time, EAC changes from negative values on Heat to $0.4757$ for QO and $0.4423$ for dense on Burgers. Distance-awareness, error-awareness, and coverage therefore capture distinct properties of the transferred uncertainty. Dense attains lower mean RMSE, MNLL, and coverage error, whereas QO has the tighter Burgers geometry bound and higher mean DAC.

Seed-level paired differences are provided in the supplementary material; on Burgers, QO has higher DAC and lower $\chi_Z$ in all three pairs, whereas dense has lower RMSE and coverage error in two of three pairs.

\paragraph{\textbf{Contextual UQ baselines}}
For broader comparison, the supplementary material reports archived in-domain MC Dropout, Deep Ensemble , and SWAG (Ref.~\cite{gal2016dropout, lakshminarayanan2017simple, maddox2019simple}) results from the earlier experimental protocol. Because those runs do not share the frozen covariance-hyperparameter transfer from Heat to Burgers, model-selection, or architectural-control protocol used here, they are treated as contextual baselines rather than matched controls. Table~\ref{tab:uq_summary} therefore remains the primary experiment for isolating QO from its dense counterpart.

\paragraph{\textbf{Post-Hoc Covariance Sensitivity}}
\label{sec:covariance_sensitivity}

To isolate covariance scale from representation learning, we hold every Burgers checkpoint, predictive mean, latent representation, RFF realization, and supervised design fixed while varying only $(\lambda,\sigma_y^2)$. Relative to the transferred setting, reducing $\lambda$ from $10^8$ to $10^2$ raises PICP2 to $0.783$ for QO and $0.872$ for dense, while increasing $\sigma_y^2$ to $10^{-2}$ at fixed $\lambda=10^8$ raises PICP2 to $0.916$ and $0.968$, respectively. The full sweep satisfies Theorem~\ref{thm:covariance_scale_new} and is reported in the supplementary material. Thus the same deterministic representation supports widely different uncertainty scales, directly confirming that representation geometry alone does not determine calibration.

\paragraph{\textbf{Finite-Readout Validation}}
\label{sec:finite_readout_exp}

We evaluate Theorem~\ref{thm:finite_readout_new} by injecting bounded signed directional perturbations satisfying $\|e\|_\infty\leq\epsilon$ into the QO checkpoints over a fixed stratified pair set and tolerance grid. Across $2{,}027{,}520$ pair/repetition evaluations, all additive-distortion, lower, and upper bounds are satisfied. The largest observed distance distortion reaches $20.9\%$ of the analytic envelope $2C_Z\epsilon$, indicating that the sufficient bound is conservative over the tested perturbations.

\begin{table}[t]
\centering
\caption{Finite-readout certificate constants for the frozen QO checkpoints. Values are mean $\pm$ population standard deviation. The sufficient tolerance $\epsilon_{\max}$ uses $\zeta=0.5$.}
\label{tab:finite_readout_validation}
\resizebox{\columnwidth}{!}{%
\begin{tabular}{l c c c c}
\hline
PDE & $C_Z$ & $r_{\rm read}/\epsilon$ & $\epsilon_{\max}(\Delta=0.05)$ & $\epsilon_{\max}(\Delta=0.10)$ \\
\hline
Heat & $27.139\pm0.653$ & $162.43\pm7.88$ & $(1.543\pm0.072)\times10^{-4}$ & $(3.085\pm0.145)\times10^{-4}$ \\
Burgers & $42.384\pm5.856$ & $398.13\pm139.80$ & $(7.006\pm2.096)\times10^{-5}$ & $(1.401\pm0.419)\times10^{-4}$ \\
\hline
\end{tabular}}
\end{table}

Burgers requires tighter readout than Heat for the same certified physical resolution because its checkpoints combine smaller $m_Z$ with larger $C_Z$. To expose the resulting scale without asserting a hardware shot count, define $R_{\rm tomo}:=\epsilon_{\max}^{-2}\log(Qd/\eta)$. With $Q=6$, $d=35$, and $\eta=0.05$, Table~\ref{tab:finite_readout_validation} gives $R_{\rm tomo}\approx3.50\times10^8$ (Heat) and $1.70\times10^9$ (Burgers) for $\Delta=0.05$, decreasing to $8.77\times10^7$ and $4.25\times10^8$ for $\Delta=0.10$. These are asymptotic access-model repetition proxies rather than measured hardware shot counts. At the studied width $d=35$, the reported certified tolerances satisfy $\epsilon_{\max}^{-2}\gg d$ for both PDEs and both resolution scales; thus the sufficient fixed-resolution bound does not establish a favorable sequential access-cost comparison at that width. This finite-width observation neither determines asymptotic scaling across widths nor provides a lower bound on the actual quantum measurement cost. The complete perturbation protocol is provided in the supplementary material.

\paragraph{\textbf{Exact Spectral-Normalization Cost}}
\label{sec:exact_svd_cost}

Each matched dense Burgers model contains six $35\times35$ spectrally normalized hidden maps and therefore performs six differentiable exact SVD normalizations per objective or closure. This yields approximately $1.89\times10^4$ normalization SVDs during training per seed and more than $3.26\times10^4$ when validation and diagnostics are included. The final effective spectral norms differ from unity by at most $1.9\times10^{-8}$. QO requires no spectral-normalization factorization because $\|W_{j,\ell}\|_2=1$ by construction. Exact per-seed counts are reported in the supplementary material.

\paragraph{\textbf{Sensitivity to Approximate Spectral Normalization}}
\label{sec:approx_sn_exp}

To isolate spectral-scale estimation error, we keep the matched-dense checkpoints fixed and replace the exact normalization scale post hoc by cold-start power-iteration estimates. Every tested realization retains a positive contraction margin and satisfies its propagated representation bounds. The effect can nevertheless be large: on Burgers with one power iteration, the minimum margin falls to $0.0099$ and the median certificate ratio increases to $\widetilde\chi_Z/\chi_Z=2.94$, whereas additional iterations rapidly recover the exact-SVD certificate. The full iteration grid, 16 estimator initializations, and all certificate statistics are reported in the supplementary material. Because weights are frozen, this experiment isolates spectral-scale estimation error rather than retraining adaptation.

Taken together, the resource results show that the two implementations incur different costs for enforcing and using the same unit-norm structure. QO avoids the repeated spectral-normalization step required by the matched dense control, but finite readout introduces a representation-dependent precision requirement on the quantum execution path. The detailed accounting and additional baseline information are provided in the supplementary material.

\section{Discussion}
\label{sec:discussion}

The forward experiments suggest that QO-SPINN is best understood as a structured hidden-layer construction rather than as a mechanism for uniformly improving approximation accuracy. Norm restoration makes the unary quantum interface faithful to the intended orthogonal map, while the pyramidal parametrization supplies both a quantum execution path and exact operator-norm control. That structural restriction remains competitive with matched dense-FiLM, and its effect on approximation is problem dependent: QO performs better on 2D advection--diffusion and similarly on the 3D problem, whereas dense-FiLM performs better on Burgers and 1D advection--diffusion. The value of the construction therefore lies primarily in the structure it imposes and exposes, not in a universal accuracy gain. The forward QO-FiLM model also inherits SPINN's rank-$r$ coordinate-separable ansatz: solutions with strong cross-coordinate coupling may require larger $r$, increasing the number of output factors and their assembly cost. This differs from reducing the solution space using POD bases or autoencoder latents, and from normalized reduced-order PINNs that reformulate the PDE as a lower-order system~\cite{luong_le-duc_lee_lee_2024}. The norm-controlled UQ model instead concatenates coordinate features and does not use a CP prediction head.

The norm-controlled experiments expose the structural benefit more directly. Exact QO layers satisfy $\|W_{j,\ell}\|_2=1$ by construction, which leads to the explicit representation and latent-kernel certificates derived in Section~\ref{sec:theory}. Exact dense spectral normalization satisfies the same sufficient operator-norm condition, but requires repeated normalization of unconstrained matrices; coarse spectral estimates can substantially loosen the resulting certificate. Finite quantum readout introduces a second source of distortion. The derived constants connect this error to a resolvable physical-coordinate scale, and the experiments confirm that the required precision depends on the trained representation: Burgers requires finer readout than Heat because it combines smaller $m_Z$ with larger $C_Z$.

The UQ experiments separate representation structure from covariance scale. The post-hoc covariance reuses the trained representation and coefficient vector, so uncertainty is obtained without an additional feature extractor or ensemble. Under covariance-hyperparameter transfer from Heat to Burgers, DAC remains high while nominal coverage collapses, and varying $(\lambda,\sigma_y^2)$ changes interval scale substantially without altering the representation. The experiments therefore support a clear division of roles: the bi-Lipschitz certificate controls latent geometry, whereas probabilistic calibration additionally depends on the learned mean and covariance hyperparameters.

Several limitations point directly to useful extensions. The expressive forward model and the certifiable residual model are currently separate realizations of the same QO primitive, so an important next step is to combine their predictive and geometric properties within a single architecture. The quantum execution analysis also remains model based: all reported predictions use classical simulation together with an explicit signed-readout access model. The fixed-$\epsilon$ access-cost analysis establishes a conditional asymptotic separation, but the sufficient precision for a fixed physical-coordinate resolution may shrink with width. In particular, the current worst-case certificate does not yield subquadratic fixed-resolution scaling for width-indexed models with nonvanishing residual scales. Assessing a practical crossover will require trained-model scaling studies or tighter readout certificates, together with hardware-level accounting for state preparation, signed measurement, total gate counts, latency, and noise. Finally, the theoretical bounds are sufficient and can be conservative, while the finite benchmark and seed set limit conclusions about predictive ordering.

\section{Conclusion}
\label{sec:conc}

We introduced QO-SPINN as a way to place quantum-compatible orthogonal transformations inside separable physics-informed models without losing the magnitude discarded by unary state preparation. Retaining the hidden-state norm classically makes the interface exact under ideal readout, while the pyramidal parametrization provides structural unit operator norm and a $\Theta(d)$-depth quantum execution path. The resulting restriction does not produce a universal accuracy advantage: QO-FiLM and matched dense-FiLM perform differently across the forward PDE benchmarks. Its main benefit is structural. In the norm-controlled residual model, exact orthogonality yields explicit bi-Lipschitz and latent-kernel bounds, finite readout can be related to a resolvable physical-coordinate scale, and the same frozen representation supports post-hoc Bayesian-linear uncertainty. The Heat-to-Burgers experiments further show that a representation can retain strong distance-aware uncertainty even when absolute calibration fails to transfer. 
% Under the signed-readout model, a width-$d$ QO map has $\widetilde{\mathcal O}(d/\epsilon^2)$ sequential access cost compared with $\mathcal O(d^2)$ dense multiplication, but the required $\epsilon$ depends on the trained representation. The present results therefore establish QO-SPINN as a structured, quantum-compatible neural primitive with explicit geometric and computational properties; determining whether those properties translate into a practical hardware advantage, once constant factors, state-preparation overheads, and hardware noise are included, remains an experimental question.
% 
Under the signed-readout model and fixed per-call failure probability, a width-$d$ QO map has sequential access-cost upper bound $\mathcal O(d\log d/\epsilon^2)$ compared with $\Theta(d^2)$ for conventional direct dense multiplication. This upper bound is asymptotically smaller when $\epsilon(d)^{-2}\log d=o(d)$; for fixed $\epsilon$, the nominal separation is nearly linear in $d$, up to logarithmic factors. In the norm-controlled residual model, the sufficient readout precision for resolving a prescribed physical-coordinate scale is determined by the trained constants $m_Z$ and $C_Z$, providing a checkpoint-level diagnostic for the width-dependent condition rather than assuming fixed precision. The present results therefore establish QO-SPINN as a structured, quantum-compatible neural primitive with explicit geometric and computational properties and a testable conditional criterion for an asymptotic hidden-map access-cost advantage; whether favorable fixed-resolution scaling can be established or translates into a practical hardware advantage, once constant factors, state-preparation costs, total gate work, signed readout, and hardware noise are included, remains an open question.

\bibliographystyle{IEEEtran}
\bibliography{orthobib}

\end{document}

\title{Supplementary Material for ``Quantum Orthogonal Separable Physics-Informed Neural Networks''}

\author{
Pietro Zanotta,
Ljubomir Budinski,
Çağlar Aytekin,
and Valtteri Lahtinen
}

\maketitle

\section{Proofs of the Representation and Uncertainty Results}
\label{sup:proofs}

This supplement provides the proofs and reproducibility details supporting the results in the main manuscript. We use the same notation as in the main paper. For each normalized physical coordinate $\bar x_j\in[0,1]$, the coordinate embedding is
\begin{equation}
E_j(\bar x_j)=
\left[
\kappa_j\bar x_j
\;\middle\|\;
\psi_j(\bar x_j)
\;\middle\|\;
0
\right]\in\R^d,
\end{equation}
where $\|\psi_j(\bar x_j)\|_2=1$. Each norm-controlled residual block is
\begin{equation}
T_{j,\ell}(h)
=
h+\rho_{j,\ell}
\sigma(W_{j,\ell}h+b_{j,\ell}),
\end{equation}
under exact readout, where $W_{j,\ell}\in SO(d)$ for QO layers.

\subsection{Coordinate-Embedding Bound}
\label{sup:coordinate_embedding}

Define
\begin{equation}
L_{\psi_j}
:=
\sup_{\bar x\in[0,1]}
\|\psi_j'(\bar x)\|_2.
\end{equation}

\begin{lemma}
For every $\bar x,\bar y\in[0,1]$,
\begin{equation}
\kappa_j|\bar x-\bar y|
\leq
\|E_j(\bar x)-E_j(\bar y)\|_2
\leq
\sqrt{\kappa_j^2+L_{\psi_j}^2}\,
|\bar x-\bar y|.
\label{sup:eq:embedding_bound}
\end{equation}
\end{lemma}

\begin{proof}
From the definition of $E_j$,
\begin{equation}
\|E_j(\bar x)-E_j(\bar y)\|_2^2
=
\kappa_j^2|\bar x-\bar y|^2
+
\|\psi_j(\bar x)-\psi_j(\bar y)\|_2^2.
\end{equation}
Nonnegativity of the second term gives the lower bound. The vector-valued mean-value inequality gives
$\|\psi_j(\bar x)-~\psi_j(\bar y)\|_2
\leq~
L_{\psi_j}|\bar x-\bar y|$,
so
\begin{equation}
\|E_j(\bar x)-E_j(\bar y)\|_2^2
\leq
(\kappa_j^2+L_{\psi_j}^2)|\bar x-\bar y|^2.
\end{equation}
Taking square roots proves \eqref{sup:eq:embedding_bound}.
\end{proof}

\subsection{Two-Sided Residual-Block Bound}
\label{sup:residual_bound}

For a residual block define
\begin{equation}
\alpha_{j,\ell}
:=
|\rho_{j,\ell}|L_\sigma
\|W_{j,\ell}\|_2.
\end{equation}

\begin{lemma}
If $\alpha_{j,\ell}<1$, then for all $h,h'\in\R^d$,
\begin{equation}
\begin{aligned}
(1-\alpha_{j,\ell})\|h-h'\|_2
&\leq \|T_{j,\ell}(h)-T_{j,\ell}(h')\|_2 \\
&\leq (1+\alpha_{j,\ell})\|h-h'\|_2.
\end{aligned}
\label{sup:eq:residual_two_sided}
\end{equation}
Moreover, $T_{j,\ell}$ is bijective and $T_{j,\ell}^{-1}$ is $(1-\alpha_{j,\ell})^{-1}$-Lipschitz.
\end{lemma}

\begin{proof}
Let
\begin{equation}
g_{j,\ell}(h)
:=
\rho_{j,\ell}
\sigma(W_{j,\ell}h+b_{j,\ell}),
\end{equation}
so that $T_{j,\ell}=I+g_{j,\ell}$. Since $\sigma$ is $L_\sigma$-Lipschitz,
\begin{equation}
\|g_{j,\ell}(h)-g_{j,\ell}(h')\|_2
\leq
\alpha_{j,\ell}\|h-h'\|_2.
\end{equation}
The triangle and reverse-triangle inequalities yield \eqref{sup:eq:residual_two_sided}. The lower bound implies injectivity when $\alpha_{j,\ell}<1$. For any $y\in\R^d$, the map
$\mathcal F_y(h):=y-g_{j,\ell}(h)$
is an $\alpha_{j,\ell}$-contraction. Banach's fixed-point theorem therefore gives a unique $h^\star$ satisfying $T_{j,\ell}(h^\star)=y$, proving surjectivity. Applying the lower bound to inverse images gives
\begin{equation}
\|T_{j,\ell}^{-1}(u)-T_{j,\ell}^{-1}(v)\|_2
\leq
(1-\alpha_{j,\ell})^{-1}
\|u-v\|_2.
\end{equation}
\end{proof}

\subsection{Complete Representation Geometry}
\label{sup:complete_geometry}

For coordinate $j$, define
\begin{equation}
m_{\Phi_j}
:=
\prod_{\ell=1}^{L_j}(1-\alpha_{j,\ell}),
\qquad
M_{\Phi_j}
:=
\prod_{\ell=1}^{L_j}(1+\alpha_{j,\ell}),
\end{equation}
and
\begin{equation}
m_j:=\kappa_jm_{\Phi_j},
\qquad
M_j:=
\sqrt{\kappa_j^2+L_{\psi_j}^2}\,M_{\Phi_j}.
\end{equation}

Let
\begin{equation}
z_j(\bar x_j)
=
\Phi_j(E_j(\bar x_j)),
\qquad
Z(\bar{\mathbf x})
=
[z_1(\bar x_1)\|\cdots\|z_K(\bar x_K)].
\end{equation}

\begin{theorem}
If every residual branch satisfies $\alpha_{j,\ell}<1$, then
\begin{equation}
m_j|\bar x_j-\bar y_j|
\leq
\|z_j(\bar x_j)-z_j(\bar y_j)\|_2
\leq
M_j|\bar x_j-\bar y_j|.
\label{sup:eq:coordinate_geometry}
\end{equation}
Furthermore,
\begin{equation}
m_Z\|\bar{\mathbf x}-\bar{\mathbf y}\|_2
\leq
\|Z(\bar{\mathbf x})-Z(\bar{\mathbf y})\|_2
\leq
M_Z\|\bar{\mathbf x}-\bar{\mathbf y}\|_2,
\label{sup:eq:global_geometry}
\end{equation}
where
\begin{equation}
m_Z:=\min_jm_j>0,
\qquad
M_Z:=\max_jM_j.
\end{equation}
Hence $Z$ is bi-Lipschitz onto its image.
\end{theorem}

\begin{proof}
Successively applying Lemma~2 gives
\begin{equation}
m_{\Phi_j}\|h-h'\|_2
\leq
\|\Phi_j(h)-\Phi_j(h')\|_2
\leq
M_{\Phi_j}\|h-h'\|_2.
\end{equation}
Combining this with Lemma~1 yields \eqref{sup:eq:coordinate_geometry}. Since
\begin{equation}
\|Z(\bar{\mathbf x})-Z(\bar{\mathbf y})\|_2^2
=
\sum_{j=1}^{K}
\|z_j(\bar x_j)-z_j(\bar y_j)\|_2^2,
\end{equation}
we obtain
\begin{equation}
m_Z^2
\|\bar{\mathbf x}-\bar{\mathbf y}\|_2^2
\leq
\|Z(\bar{\mathbf x})-Z(\bar{\mathbf y})\|_2^2
\leq
M_Z^2
\|\bar{\mathbf x}-\bar{\mathbf y}\|_2^2.
\end{equation}
Taking square roots proves \eqref{sup:eq:global_geometry}.
\end{proof}

\subsection{Inexact Dense Spectral Normalization}
\label{sup:inexact_sn}

For a dense matrix $A_{j,\ell}$ normalized using an estimated spectral scale $\widehat s_{j,\ell}>0$, let
\begin{equation}
\widetilde W_{j,\ell}
=
\frac{A_{j,\ell}}{\widehat s_{j,\ell}},
\qquad
\tau_{j,\ell}
=
\frac{\sigma_{\max}(A_{j,\ell})}
{\widehat s_{j,\ell}}
=
\|\widetilde W_{j,\ell}\|_2.
\end{equation}

\begin{corollary}
If $\tau_{j,\ell}\leq1+\xi_{j,\ell}$, then the corresponding residual branch has Lipschitz constant at most
\begin{equation}
\bar\alpha_{j,\ell}
=
|\rho_{j,\ell}|L_\sigma(1+\xi_{j,\ell}).
\end{equation}
Whenever $\bar\alpha_{j,\ell}<1$ for all blocks, the representation bounds above hold after replacing $\alpha_{j,\ell}$ by $\bar\alpha_{j,\ell}$.
\end{corollary}

\begin{proof}
By construction,
\begin{equation}
\|\widetilde W_{j,\ell}\|_2
=
\tau_{j,\ell}
\leq
1+\xi_{j,\ell}.
\end{equation}
Therefore
\begin{equation}
\operatorname{Lip}
\left(
\rho_{j,\ell}
\sigma(
\widetilde W_{j,\ell}\,\cdot+b_{j,\ell})
\right)
\leq
|\rho_{j,\ell}|L_\sigma(1+\xi_{j,\ell})
=
\bar\alpha_{j,\ell}.
\end{equation}
The result follows by applying the residual-block argument with $\bar\alpha_{j,\ell}$ and composing the resulting layer-wise constants.
\end{proof}

\subsection{Latent-Kernel Consequence}
\label{sup:kernel_geometry}

Consider
\begin{equation}
k_Z(\mathbf x,\mathbf y)
=
\exp
\left[
-\gamma
\|Z(\bar{\mathbf x})-Z(\bar{\mathbf y})\|_2^2
\right].
\end{equation}

\begin{corollary}
Under the assumptions of Theorem~1,
\begin{equation}
\begin{aligned}
\exp\left[-\gamma M_Z^2\|\bar{\mathbf x}-\bar{\mathbf y}\|_2^2\right]
&\leq k_Z(\mathbf x,\mathbf y) \\
&\leq \exp\left[-\gamma m_Z^2\|\bar{\mathbf x}-\bar{\mathbf y}\|_2^2\right].
\end{aligned}
\label{sup:eq:kernel_bounds}
\end{equation}
\end{corollary}

\begin{proof}
Theorem~1 bounds the squared latent distance between
$m_Z^2\|\bar{\mathbf x}-\bar{\mathbf y}\|_2^2$
and
$M_Z^2\|\bar{\mathbf x}-\bar{\mathbf y}\|_2^2$.
Multiplying by $-\gamma$ reverses the order, and monotonicity of the exponential yields \eqref{sup:eq:kernel_bounds}.
\end{proof}

Because $Z$ is formed by concatenation,
\begin{equation}
k_Z(\mathbf x,\mathbf y)
=
\prod_{j=1}^{K}
\exp
\left[
-\gamma
\|z_j(\bar x_j)-z_j(\bar y_j)\|_2^2
\right].
\end{equation}

\subsection{Finite-Readout Representation Error}
\label{sup:finite_readout}

Assume
\begin{equation}
|\sigma(a)_i|
\leq
S_\sigma.
\end{equation}
For the sine activation used in the experiments, $S_\sigma=1$.

Define
\begin{equation}
B_{j,0}
:=
\sqrt{\kappa_j^2+1},
\qquad
B_{j,\ell}
:=
B_{j,0}
+
S_\sigma\sqrt d
\sum_{r=1}^{\ell}
|\rho_{j,r}|.
\end{equation}

If the directional reconstruction at layer $(j,\ell)$ satisfies a componentwise error bound $\epsilon_{j,\ell}$, define
\begin{equation}
\delta_{j,\ell}
:=
|\rho_{j,\ell}|L_\sigma
B_{j,\ell-1}
\sqrt d\,
\epsilon_{j,\ell},
\end{equation}
\begin{equation}
D_j
:=
\sum_{\ell=1}^{L_j}
\delta_{j,\ell}
\prod_{r=\ell+1}^{L_j}
(1+\alpha_{j,r}),
\qquad
D_Z
:=
\left(
\sum_{j=1}^{K}D_j^2
\right)^{1/2}.
\end{equation}

\begin{theorem}
Let
\begin{equation}
Q:=\sum_{j=1}^{K}L_j
\end{equation}
be the number of QO readout calls required for one complete feature-extractor evaluation. For a fixed pair $(\bar{\mathbf x},\bar{\mathbf y})$, suppose each of the corresponding $2Q$ calls satisfies its componentwise readout event with conditional failure probability at most $\eta/(2Q)$. Then, with probability at least $1-\eta$,
\begin{equation}
m_Z
\|\bar{\mathbf x}-\bar{\mathbf y}\|_2
-
2D_Z
\leq
\|\widehat Z(\bar{\mathbf x})-\widehat Z(\bar{\mathbf y})\|_2
\leq
M_Z
\|\bar{\mathbf x}-\bar{\mathbf y}\|_2
+
2D_Z.
\label{sup:eq:finite_readout_geometry}
\end{equation}
\end{theorem}

\begin{proof}
Since $\|\psi_j(\bar x_j)\|_2=1$ and $\bar x_j\in[0,1]$,
\begin{equation}
\|\widehat h_{j,0}\|_2
=
\|E_j(\bar x_j)\|_2
\leq
B_{j,0}.
\end{equation}
Moreover,
\begin{equation}
\|\sigma(a)\|_2
\leq
S_\sigma\sqrt d.
\end{equation}
Induction through the noisy residual recursion therefore gives
\begin{equation}
\|\widehat h_{j,\ell}\|_2
\leq
B_{j,\ell}.
\end{equation}

Let
\begin{equation}
e_{j,\ell}
:=
\|\widehat h_{j,\ell}-h_{j,\ell}\|_2.
\end{equation}
On a successful readout event,
\begin{align}
e_{j,\ell}
&\leq
e_{j,\ell-1}
+
|\rho_{j,\ell}|L_\sigma
\|
\widehat{\mathcal R}_{W_{j,\ell}}
(\widehat h_{j,\ell-1})
-
W_{j,\ell}h_{j,\ell-1}
\|_2
\nonumber\\
&\leq
(1+\alpha_{j,\ell})e_{j,\ell-1}
+
\delta_{j,\ell}.
\end{align}
Because $e_{j,0}=0$, iteration yields
\begin{equation}
e_{j,L_j}
\leq
\sum_{\ell=1}^{L_j}
\delta_{j,\ell}
\prod_{r=\ell+1}^{L_j}
(1+\alpha_{j,r})
=
D_j.
\end{equation}
Concatenation then gives
\begin{equation}
\|\widehat Z(\bar{\mathbf x})-Z(\bar{\mathbf x})\|_2
\leq
D_Z.
\end{equation}
Applying the same bound to $\bar{\mathbf y}$ gives
\begin{equation}
\left|
\|\widehat Z(\bar{\mathbf x})-\widehat Z(\bar{\mathbf y})\|_2
-
\|Z(\bar{\mathbf x})-Z(\bar{\mathbf y})\|_2
\right|
\leq
2D_Z.
\end{equation}
Combining this with Theorem~1 proves \eqref{sup:eq:finite_readout_geometry}. A union bound over the $2Q$ conditional readout events gives joint success probability at least $1-\eta$.
\end{proof}

\subsection{Physical-Resolution Requirement}
\label{sup:resolution}

For a common tolerance $\epsilon_{j,\ell}=\epsilon$, write
\begin{equation}
D_Z=C_Z\epsilon,
\end{equation}
where
\begin{equation}
\begin{aligned}
C_j
&:=
\sum_{\ell=1}^{L_j}
|\rho_{j,\ell}|L_\sigma
B_{j,\ell-1}\sqrt d
\prod_{r=\ell+1}^{L_j}
(1+\alpha_{j,r}), \\
C_Z
&:=
\left(
\sum_{j=1}^{K}C_j^2
\right)^{1/2}.
\end{aligned}
\end{equation}

\begin{corollary}
Fix a nondimensionalized-coordinate separation $\Delta>0$ and $\zeta\in(0,1)$. If
\begin{equation}
\epsilon
\leq
\frac{(1-\zeta)m_Z\Delta}{2C_Z},
\label{sup:eq:epsilon_resolution}
\end{equation}
then every fixed queried pair satisfying
$\|\bar{\mathbf x}-\bar{\mathbf y}\|_2\geq\Delta$
obeys
\begin{equation}
\|\widehat Z(\bar{\mathbf x})-
\widehat Z(\bar{\mathbf y})\|_2
\geq
\zeta m_Z\Delta
\end{equation}
on the corresponding joint readout-success event.
\end{corollary}

\begin{proof}
Theorem~2 gives
\begin{equation}
\|\widehat Z(\bar{\mathbf x})-
\widehat Z(\bar{\mathbf y})\|_2
\geq
m_Z\Delta-2C_Z\epsilon.
\end{equation}
Substituting \eqref{sup:eq:epsilon_resolution} yields
$\zeta m_Z\Delta$.
\end{proof}

Under the signed $\ell_\infty$ access model,
\begin{equation}
N_{\rm tomo}
=
\mathcal O
\left(
\frac{\log(Qd/\eta)}{\epsilon^2}
\right),
\end{equation}
so a sufficient repetition budget is
\begin{equation}
N_{\rm tomo}
=
\mathcal O
\left(
\frac{C_Z^2}
{(1-\zeta)^2m_Z^2\Delta^2}
\log\frac{Qd}{\eta}
\right),
\end{equation}
up to access-model constants. A concrete instantiation of these constants for the signed $\ell_\infty$ tomography construction of \cite{kerenidis2019quantum} is given in Section~\ref{sup:readout_scale}. Hence the sufficient tomography budget scales as $\Delta^{-2}$.

\subsection{Composition With the Hidden-Map Complexity Bound}
\label{sup:fixed_resolution_complexity}

For $H$ hidden-map calls, the signed-readout access model gives
\begin{equation}
T_{\rm hidden}^{\rm QO,quantum}
=
\mathcal O
\left(
\frac{Hd\log(dH/\eta)}
{\epsilon^2}
\right).
\end{equation}
Combining this expression with \eqref{sup:eq:epsilon_resolution} gives the sufficient fixed-resolution access cost
\begin{equation}
T_{\rm hidden,\Delta}^{\rm QO,quantum}
=
\mathcal O
\left(
\frac{
HdC_Z^2
}{
(1-\zeta)^2m_Z^2\Delta^2
}
\log\frac{dH}{\eta}
\right),
\label{sup:eq:fixed_resolution_complexity}
\end{equation}
where constant numerical factors are absorbed by $\mathcal O(\cdot)$.

Equation~\eqref{sup:eq:fixed_resolution_complexity} is deliberately a sufficient access-model bound, not a hardware-runtime prediction. In particular, $C_Z$ and $m_Z$ depend on the trained representation and may themselves vary with width. Therefore the fixed-$\epsilon$ scaling
$\widetilde{\mathcal O}(Hd/\epsilon^2)$
does not imply a universal subquadratic guarantee when the certified physical resolution $\Delta$ is held fixed.

\subsection{Post-Hoc Covariance Scale}
\label{sup:covariance_scale}

Let
\begin{equation}
G
=
\Phi_y^T\Phi_y
\succeq0,
\qquad
\tau
=
\sigma_y^2,
\qquad
\Sigma(\lambda,\tau)
=
(\lambda I+\tau^{-1}G)^{-1}.
\end{equation}

\begin{theorem}
For $\lambda,\tau>0$,
\begin{equation}
0
\prec
\Sigma(\lambda,\tau)
\preceq
\lambda^{-1}I.
\end{equation}
Consequently,
\begin{equation}
0
\leq
\sigma_{\rm epi,*}^2
\leq
\frac{\|\varphi_*\|_2^2}{\lambda}.
\end{equation}
Moreover,
\begin{equation}
\frac{\partial\Sigma}{\partial\lambda}
=
-\Sigma^2
\preceq0,
\qquad
\frac{\partial\Sigma}{\partial\tau}
=
\tau^{-2}\Sigma G\Sigma
\succeq0.
\end{equation}
Thus epistemic variance decreases with $\lambda$ and increases with $\tau$, while
$\sigma_{\rm pred,*}^2=\sigma_{\rm epi,*}^2+\tau$
increases strictly with $\tau$.
\end{theorem}

\begin{proof}
Because
\begin{equation}
\lambda I+\tau^{-1}G
\succeq
\lambda I
\succ0,
\end{equation}
matrix inversion reverses the Loewner order:
\begin{equation}
0
\prec
\Sigma(\lambda,\tau)
\preceq
\lambda^{-1}I.
\end{equation}
Taking the quadratic form with $\varphi_*$ gives the variance bound.

For a differentiable positive-definite matrix $A(\theta)$,
\begin{equation}
\frac{\partial A^{-1}}{\partial\theta}
=
-A^{-1}
\frac{\partial A}{\partial\theta}
A^{-1}.
\end{equation}
Applying this identity to
$A=\lambda I+\tau^{-1}G$
yields
\begin{equation}
\frac{\partial\Sigma}{\partial\lambda}
=
-\Sigma^2,
\qquad
\frac{\partial\Sigma}{\partial\tau}
=
\tau^{-2}\Sigma G\Sigma.
\end{equation}
Taking quadratic forms gives
\begin{equation}
\frac{\partial\sigma_{\rm epi,*}^2}
{\partial\lambda}
=
-\|\Sigma\varphi_*\|_2^2
\leq0,
\end{equation}
and
\begin{equation}
\frac{\partial\sigma_{\rm epi,*}^2}
{\partial\tau}
=
\tau^{-2}
\varphi_*^T\Sigma G\Sigma\varphi_*
\geq0.
\end{equation}
Finally,
\begin{equation}
\frac{\partial\sigma_{\rm pred,*}^2}
{\partial\tau}
=
1+
\tau^{-2}
\varphi_*^T\Sigma G\Sigma\varphi_*
>
0.
\end{equation}
\end{proof}

\subsection{Representation Geometry Does Not Determine Calibration}
\label{sup:geometry_calibration}

\begin{corollary}
The geometric constants $m_Z$, $M_Z$, and their ratio do not determine a positive lower bound on epistemic variance or nominal predictive coverage.
\end{corollary}

\begin{proof}
Fix the deterministic representation $Z$, random-feature realization, $G=\Phi_y^T\Phi_y$, and a test feature $\varphi_*$. The geometric constants are independent of $\lambda$, whereas
\begin{equation}
\Sigma(\lambda,\tau)
=
\lambda^{-1}
\left(
I+\frac{G}{\lambda\tau}
\right)^{-1}
\longrightarrow0
\end{equation}
as $\lambda\rightarrow\infty$. Therefore
\begin{equation}
\sigma_{\rm epi,*}^2
=
\varphi_*^T
\Sigma(\lambda,\tau)
\varphi_*
\longrightarrow0
\end{equation}
without changing the deterministic representation geometry. Thus representation geometry alone cannot determine the scale of uncertainty or predictive coverage.
\end{proof}

\section{Experimental and Reproducibility Details}
\label{sup:reproducibility}

\subsection{Forward QO-FiLM Experiments}
\label{sup:forward_details}

The forward QO-FiLM and matched dense-FiLM experiments use paired seeds
\begin{equation}
\{42,123,456\}.
\end{equation}
Both models use width $d=20$, four Fourier frequencies, and $20{,}000$ full-batch Adam updates. Within each QO--dense pair, the data, width, depth, Fourier initialization, non-map tensors, optimization protocol, and initial numerical square map are shared.

Burgers uses CP rank $32$ and learning rate $5\times10^{-3}$. The one-, two-, and three-dimensional advection--diffusion experiments use CP rank $16$ and learning rate $10^{-3}$.

Checkpoint selection is restricted to updates $18{,}001$--$20{,}000$ and is applied identically within each matched QO--dense pair. Final evaluation is performed on deterministic cell-midpoint grids that are not used for checkpoint selection. The evaluation sizes are $99\times99$ for Burgers and one-dimensional advection--diffusion, $39^3$ for two-dimensional advection--diffusion, and $11^4$ for three-dimensional advection--diffusion. Analytical reference solutions are evaluated directly at the final evaluation points without interpolation.

\begin{table}[t]
\centering
\caption{Fixed forward-regression setup. QO-FiLM and dense-FiLM use two width-$d$ square hidden maps per coordinate subnetwork, with the CP output head counted separately.}\label{sup:tab:forward_setup}
\small
\setlength{\tabcolsep}{2pt}
\begin{tabular*}{\columnwidth}{@{\extracolsep{\fill}}l c c c c}
\hline
Problem & Width & CP rank & Adam LR & Evaluation grid \\
\hline
Burgers & 20 & 32 & $5\times10^{-3}$ & $99\times99$ \\
Adv.--Diff. 1D & 20 & 16 & $10^{-3}$ & $99\times99$ \\
Adv.--Diff. 2D & 20 & 16 & $10^{-3}$ & $39^3$ \\
Adv.--Diff. 3D & 20 & 16 & $10^{-3}$ & $11^4$ \\
\hline
\end{tabular*}
\end{table}

\begin{table*}[t]
\centering
\caption{Contextual published forward benchmarks for the PDE tasks used in the main forward study. Entries are mean $\pm$ population standard deviation as reported for the corresponding evaluations. These external baselines provide task-level context and are not matched architectural controls for QO-FiLM.}
\label{sup:tab:contextual_forward_results}
\small
\setlength{\tabcolsep}{4pt}
\renewcommand{\arraystretch}{1.18}
\begin{tabular*}{\textwidth}{@{\extracolsep{\fill}}llcccc@{}}
\hline
\textbf{Equation} & \textbf{Metric} & \textbf{PINN} & \textbf{SPINN} & \textbf{XPINN} & \textbf{PIKAN} \\
\hline
\multirow{2}{*}{Burgers (1D)} & Rel. $L_2$ & \meanstd{2.74\times10^{-4}}{1.01\times10^{-5}} & \meanstd{5.95\times10^{-4}}{1.49\times10^{-5}} & \meanstd{5.13\times10^{-3}}{3.91\times10^{-4}} & \meanstd{8.83\times10^{-4}}{6.38\times10^{-4}} \\
& Max Error & \meanstd{6.01\times10^{-4}}{1.86\times10^{-4}} & \meanstd{1.49\times10^{-3}}{1.77\times10^{-4}} & \meanstd{8.69\times10^{-3}}{1.46\times10^{-3}} & \meanstd{1.02\times10^{-3}}{4.60\times10^{-4}} \\
\hline
\multirow{2}{*}{Adv.-Diff. (1D)} & Rel. $L_2$ & \meanstd{1.27\times10^{-3}}{6.25\times10^{-5}} & \meanstd{1.86\times10^{-3}}{7.58\times10^{-4}} & \meanstd{1.56\times10^{-3}}{4.41\times10^{-4}} & \meanstd{8.78\times10^{-4}}{1.63\times10^{-4}} \\
& Max Error & \meanstd{1.14\times10^{-1}}{1.17\times10^{-2}} & \meanstd{4.10\times10^{-2}}{1.22\times10^{-2}} & \meanstd{1.74\times10^{-1}}{3.41\times10^{-2}} & \meanstd{4.92\times10^{-2}}{1.30\times10^{-2}} \\
\hline
\multirow{2}{*}{Adv.-Diff. (2D)} & Rel. $L_2$ & \meanstd{7.38\times10^{-3}}{8.36\times10^{-4}} & \meanstd{2.49\times10^{-3}}{3.66\times10^{-4}} & \meanstd{9.81\times10^{-3}}{6.59\times10^{-4}} & \meanstd{2.63\times10^{-3}}{1.06\times10^{-4}} \\
& Max Error & \meanstd{2.59}{0.767} & \meanstd{0.285}{0.0413} & \meanstd{2.87}{0.376} & \meanstd{0.853}{0.142} \\
\hline
\multirow{2}{*}{Adv.-Diff. (3D)} & Rel. $L_2$ & \meanstd{0.367}{0.313} & \meanstd{0.0138}{0.00247} & \meanstd{0.406}{0.272} & \meanstd{0.158}{0.137} \\
& Max Error & \meanstd{83.8}{56.3} & \meanstd{3.25}{0.562} & \meanstd{92.1}{43.5} & \meanstd{36.6}{16.0} \\
\hline
\end{tabular*}
\end{table*}

\subsection{Norm-Controlled UQ Experiments}
\label{sup:uq_details}

The matched QO and dense UQ models use paired seeds
$\{42,123,456\}$,
width $d=35$, three residual blocks per coordinate, sine activation, the same normalized-coordinate/Fourier embedding, and the same global random-feature realization within each pair. The architectures differ only in their square hidden maps: exact RBS orthogonal transformations for QO and exact-SVD spectral normalization for dense.

The transferred Heat configuration uses $\rho_{\max}=0.5$, $\delta_\rho=0.25$, $\kappa=1$, $L_\sigma=1$, $D_L=1024$, $\gamma=5$, $\lambda=10^8$, and $\sigma_y^2=3\times10^{-8}$. The covariance parameters were selected after deterministic training with all network parameters, predictive coefficients, latent representations, and random-feature realizations held fixed. Calibration used 500 additional Latin-hypercube points on the Heat domain with analytical reference values, disjoint from the training, checkpoint-selection, and final-evaluation points. A single common pair $(\lambda,\sigma_y^2)$ was selected for both architectures by minimizing calibration MNLL averaged equally over the six Heat runs comprising QO and dense models with seeds $42$, $123$, and $456$. An initial 35-pair search attained its minimum at $(10^6,10^{-7})$; because $10^6$ was the largest tested prior precision, we subsequently evaluated the boundary-check grid $\lambda\in\{10^6,3\times10^6,10^7,3\times10^7,10^8,3\times10^8,10^9\}$ and $\sigma_y^2\in\{3\times10^{-8},10^{-7},3\times10^{-7}\}$. The minimum pooled calibration MNLL on this grid was $-7.0250101$ at $(\lambda,\sigma_y^2)=(10^8,3\times10^{-8})$. These values optimize the stated finite-grid MNLL criterion rather than representing theoretically prescribed constants or a demonstrated continuous optimum; in particular, the selected $\sigma_y^2$ lies at the lower boundary of this grid. The resulting pair was fixed before the Burgers evaluation and transferred without target-domain recalibration. Here $\delta_\rho$ is the implementation admissibility guard associated with $|\rho|\leq1-\delta_\rho$; it is distinct from the residual-scale initialization, and the chosen $\rho_{\max}=0.5$ already imposes the tighter bound $|\rho|\leq0.5$. For both coordinates, $p_j=4$, $\xi_j=(1,2,4,8)/\sqrt{85}$, and $\nu_j=(0,0,0,0)$, with no additional $2\pi$ factor in the Fourier arguments. These coordinate-frequency and phase buffers are fixed in the Heat configuration and transferred unchanged to Burgers.

Heat solves
\begin{equation}
u_t
=
\mu u_{xx},
\qquad
\mu
=
\frac{1}{\pi^2},
\end{equation}
on $[-1,1]\times[0,1]$. The supervised data contain $20$ initial-condition observations and $40$ boundary observations. The PDE residual uses $100$ fixed collocation points taken directly from the archived \texttt{collocationpoint.mat} coordinate array; the same collocation coordinates are reused across seeds and across the matched QO and dense architectures rather than regenerated. Model selection uses $500$ Latin-hypercube validation points per seed.

Burgers solves
\begin{equation}
u_t+uu_x
=
\nu u_{xx},
\qquad
\nu
=
\frac{0.1}{\pi},
\end{equation}
with
\begin{equation}
u(x,0)
=
-\sin(\pi x)
\end{equation}
and homogeneous Dirichlet boundary conditions on $[-1,1]\times[0,1]$.

The Burgers sparse-data protocol contains $20$ initial-condition locations and $20$ boundary times at each endpoint, giving $60$ supervised observations, together with $100$ PDE collocation points. Model selection uses $500$ Latin-hypercube validation points per seed. Final evaluation uses a $101\times101$ grid against the Cole--Hopf reference obtained from the archived $256\times100$ quadrature data.

For Burgers, training consists of $1000$ Adam steps with learning rate $10^{-3}$ followed by at most $100$ L-BFGS calls initialized from the best Adam validation state. The Burgers objective is
\begin{equation}
50L_{\rm IC}
+
50(L_{\rm BC,left}+L_{\rm BC,right})
+
10L_{\rm PDE}
+
10^{-6}\|\widehat\beta\|_2^2.
\end{equation}
For Heat, the coefficient penalty is $10^{-6}\|\widehat\beta\|_2^2$ during Adam and $10^{-4}\|\widehat\beta\|_2^2$ during L-BFGS, and checkpoint-selection eligibility begins in the final $10\%$ of the Adam phase. Within each PDE, the matched QO and dense models use the same optimization and model-selection protocol.

\begin{table}[t]
\centering
\caption{Norm-controlled UQ configuration shared within each QO--dense pair.}
\label{sup:tab:uq_setup}
\small
\setlength{\tabcolsep}{2pt}
\begin{tabular*}{\columnwidth}{@{\extracolsep{\fill}}l c}
\hline
Quantity & Value \\
\hline
Width $d$ & 35 \\
Residual blocks/coordinate & 3 \\
Activation & Sine \\
$\rho_{\max}$ & 0.5 \\
$\kappa$ & 1 \\
RFF dimension $D_L$ & 1024 \\
Kernel parameter $\gamma$ & 5 \\
Prior precision $\lambda$ & $10^8$ \\
Observation variance $\sigma_y^2$ & $3\times10^{-8}$ \\
Outer seeds & 42, 123, 456 \\
\hline
\end{tabular*}
\end{table}

\section{Quantum Resource Proxies}
\label{sup:resources}

The quantities below are algorithmic resource proxies and not hardware-runtime measurements. All numerical experiments in the paper evaluate QO maps through exact classical RBS simulation.

\noindent\textbf{Why RBS gates yield $SO(d)$.} On the $d$-dimensional unary subspace, an RBS gate acting on neighboring modes $(i,i+1)$ induces the real planar rotation $G(\theta)=\bigl[\begin{smallmatrix}\cos\theta&\sin\theta\\-\sin\theta&\cos\theta\end{smallmatrix}\bigr]$ on their two basis states and acts as the identity on all other unary basis states. Its embedded matrix $R_i(\theta)\in\mathbb R^{d\times d}$ therefore obeys $R_i(\theta)^TR_i(\theta)=I_d$ and $\det R_i(\theta)=\cos^2\theta+\sin^2\theta=1$. A pyramidal composition $W=R_{i_P}(\theta_P)\cdots R_{i_1}(\theta_1)$ consequently satisfies $W^TW=I_d$ and $\det W=\prod_{k=1}^P\det R_{i_k}(\theta_k)=1$, proving $W\in SO(d)$. This conclusion concerns the unary subspace, not the full $2^d$-dimensional Hilbert space.

For a square width-$d$ pyramidal orthogonal map, the trainable RBS count is
\begin{equation}
P_d
=
\frac{d(d-1)}{2},
\end{equation}
the unary state-preparation circuit requires $d-1$ RBS rotations, and the pyramid has $2d-3$ sequential rotation timesteps. Signed amplitude reconstruction additionally requires the readout procedure assumed in the main paper. Under a one-ancilla implementation, the logical-qubit proxy is $d+1$.

\begin{table}[t]
\centering
\caption{Per-map QO resource proxies for the widths used in the experiments. Qubit counts include one tomography ancilla.}
\label{sup:tab:qresources}
\small
\setlength{\tabcolsep}{2pt}
\begin{tabular*}{\columnwidth}{@{\extracolsep{\fill}}l c c c c c}
\hline
Study & $d$ & Qubits & RBS/map & RBS/load & Pyramid depth \\
\hline
Forward & 20 & 21 & 190 & 19 & 37 \\
UQ & 35 & 36 & 595 & 34 & 67 \\
\hline
\end{tabular*}
\end{table}

These counts intentionally separate total gate count from sequential depth. Although a width-$d$ pyramid contains $\Theta(d^2)$ RBS rotations, rotations acting on disjoint mode pairs execute in parallel, giving $\Theta(d)$ depth. The signed-readout repetitions are additional and are analyzed separately through the tolerance-dependent access model.

\subsection{Readout-Precision Scale}
\label{sup:readout_scale}

For the frozen QO checkpoints, the sufficient finite-readout constants are summarized below.

\begin{table*}[t]
\centering
\caption{Finite-readout certificate constants for the frozen QO checkpoints. Values are mean $\pm$ population standard deviation over seeds 42, 123, and 456. The sufficient tolerance uses $\zeta=0.5$.}
\label{sup:tab:finite_readout}
\small
\setlength{\tabcolsep}{3pt}
\begin{tabular*}{\textwidth}{@{\extracolsep{\fill}}l c c c c}
\hline
PDE & $C_Z$ & $r_{\rm read}/\epsilon$ & $\epsilon_{\max}(0.05)$ & $\epsilon_{\max}(0.10)$ \\
\hline
Heat & $27.139\pm0.653$ & $162.43\pm7.88$ & $(1.543\pm0.072)\times10^{-4}$ & $(3.085\pm0.145)\times10^{-4}$ \\
Burgers & $42.384\pm5.856$ & $398.13\pm139.80$ & $(7.006\pm2.096)\times10^{-5}$ & $(1.401\pm0.419)\times10^{-4}$ \\
\hline
\end{tabular*}
\end{table*}

For a transparent numerical scale, with $\log$ denoting the natural logarithm, define the asymptotic repetition proxy
\begin{equation}
R_{\rm tomo}
:=
\epsilon_{\max}^{-2}
\log\frac{Qd}{\eta}.
\end{equation}
For the present UQ architecture, $Q=6$ and $d=35$. Taking $\eta=0.05$, the mean tolerances in Table~\ref{sup:tab:finite_readout} give approximately
\begin{equation}
R_{\rm tomo}
\approx
3.50\times10^8
\quad\text{(Heat)},\qquad
1.70\times10^9
\quad\text{(Burgers)}
\end{equation}
for $\Delta=0.05$, and
\begin{equation}
R_{\rm tomo}
\approx
8.77\times10^7
\quad\text{(Heat)},\qquad
4.25\times10^8
\quad\text{(Burgers)}
\end{equation}
for $\Delta=0.10$.

These quantities are not asserted to be physical hardware shot counts. They expose the numerical scale of the generalized asymptotic access model after substituting the sufficient representation-dependent precision.

To make the hidden constant concrete for one signed $\ell_\infty$ tomography construction, Algorithm~3 of Ref.~\cite{kerenidis2019quantum} uses two batches of $\left\lceil 36\ln(d)/\delta^2\right\rceil$ state preparations and measurements. Theorem~G.4 of that work guarantees a final reconstruction error smaller than $(1+\sqrt{2})\delta$ with probability at least $1-d^{-0.83}$. Therefore, to guarantee a target componentwise reconstruction tolerance $\epsilon$ under that construction, we set $\delta=\epsilon/(1+\sqrt{2})$, giving
\begin{equation}
N_{\rm KLP}(\epsilon,d)=2\left\lceil 36(1+\sqrt{2})^2\frac{\ln d}{\epsilon^2}\right\rceil.
\label{sup:eq:klp_repetitions}
\end{equation}
For the UQ width $d=35$, this is approximately $1.492\times10^3\,\epsilon^{-2}$ state-preparation/measurement repetitions per signed readout. Substituting the sufficient mean tolerances from Table~\ref{sup:tab:finite_readout} gives the concrete values below.

\begin{table}[t]
\centering
\caption{Explicit per-readout repetition counts for the signed $\ell_\infty$ tomography construction of \cite{kerenidis2019quantum} at the sufficient UQ tolerances. The displayed tolerances are rounded, while the counts are evaluated using the corresponding unrounded mean tolerances underlying Table~\ref{sup:tab:finite_readout}.}
\label{sup:tab:klp_repetitions}
\small
\setlength{\tabcolsep}{4pt}
\begin{tabular*}{\columnwidth}{@{\extracolsep{\fill}}l c c c}
\hline
PDE & $\Delta$ & $\epsilon_{\max}$ & $N_{\rm KLP}$/readout \\
\hline
Heat & $0.05$ & $1.543\times10^{-4}$ & $6.269\times10^{10}$ \\
Heat & $0.10$ & $3.085\times10^{-4}$ & $1.567\times10^{10}$ \\
Burgers & $0.05$ & $7.006\times10^{-5}$ & $3.040\times10^{11}$ \\
Burgers & $0.10$ & $1.401\times10^{-4}$ & $7.600\times10^{10}$ \\
\hline
\end{tabular*}
\end{table}

For $d=35$, the native success probability of this explicit construction is $1-35^{-0.83}\approx0.9477$ per tomography call. This is distinct from the arbitrary-confidence model used in $R_{\rm tomo}$, where the per-call failure probability is allocated through $\eta$; achieving a smaller failure probability with the explicit construction requires a separate confidence-amplification procedure and therefore an additional scheme-dependent constant. The factor $d$ in the sequential-access complexity $\mathcal O(d\log d/\epsilon^2)$ arises because each tomography repetition requires a state-preparation-plus-pyramid execution. For $d=35$, Table~\ref{sup:tab:qresources} gives a pyramid depth of $67$ sequential rotation timesteps and $34$ RBS rotations for unary state preparation. Equation~\eqref{sup:eq:klp_repetitions} therefore makes the tomography repetition constant explicit, while the exact device-level constant multiplying the linear-depth circuit remains implementation dependent. These values are idealized state-preparation/measurement repetition counts under the stated access assumptions; they are not measured physical-device shot budgets or wall-clock runtimes, and they do not include device latency, parallelism, gate noise, error correction, or end-to-end training multiplicities.

\section{Geometry and Numerical Validation Protocol}
\label{sup:geometry_protocol}

For exact-geometry checks, effective operator norms and geometric constants are recomputed in \texttt{float64}. The Burgers pair set is generated independently of model outcomes using pair seed 20261002 and contains identical, very-short-distance, medium-distance, long-distance, and uniformly sampled pairs.

In the archived primary Heat and Burgers geometry checks, a sampled lower or upper latent-distance inequality is counted as violated only when its violation residual exceeds the implemented tolerance $2\times10^{-12}+2\times10^{-9}\|\bar{\mathbf x}-\bar{\mathbf y}\|_2$. Identical pairs therefore require
\begin{equation}
\|Z(\bar{\mathbf x})-Z(\bar{\mathbf y})\|_2
\leq
2\times10^{-12}.
\end{equation}
The corresponding latent-kernel checks evaluate $\exp[-\gamma\|Z(\bar{\mathbf x})-Z(\bar{\mathbf y})\|_2^2]$ directly, rather than in the log domain, and count a kernel-bound violation only when the absolute violation residual exceeds $2\times10^{-12}$.

\subsection{Finite-Readout Perturbation Protocol}
\label{sup:perturbation_protocol}

Finite-readout validation uses the six frozen QO checkpoints from Heat and Burgers (three seeds per PDE), ten tolerances $\epsilon\in\{10^{-6},3\times10^{-6},10^{-5},3\times10^{-5},10^{-4},3\times10^{-4},10^{-3},3\times10^{-3},10^{-2},3\times10^{-2}\}$, two perturbation families, $32$ repetitions, and $528$ input pairs consisting of $128$ very-short, $128$ short, $128$ medium, $128$ long, and $16$ identical pairs. The perturbations are either independent componentwise uniform errors in $[-\epsilon,\epsilon]$ or Rademacher errors in $\{-\epsilon,+\epsilon\}$ and satisfy
\begin{equation}
\|e\|_\infty
\leq
\epsilon.
\end{equation}
The perturbed readout vector is not renormalized. These perturbations test the bounded-error event assumed by the finite-readout theorem; they are not intended as a device-level tomography simulator. The resulting total is $2\times3\times10\times2\times32\times528=2{,}027{,}520$ pair/repetition evaluations. All tested additive-distortion, lower, and upper bounds are satisfied, and the largest observed distance distortion is $20.9025\%$ of the analytic envelope $2C_Z\epsilon$ (reported as $20.9\%$ in the main text), showing that the sufficient bound is conservative over the tested perturbations.

A separate multinomial diagnostic based on squared amplitudes and oracle signs is not used to validate the signed-$\ell_\infty$ access model because it represents a different measurement model.

\section{Dense Spectral-Normalization Accounting}
\label{sup:spectral_accounting}

Each matched dense Burgers model contains six $35\times35$ hidden maps normalized through differentiable exact singular-value decompositions. Seeds 42, 123, and 456 require, respectively,
\begin{equation}
18{,}852,\quad
18{,}864,\quad
18{,}894
\end{equation}
training normalization SVDs.

Including validation and diagnostic evaluations gives
\begin{equation}
32{,}676,\quad
32{,}688,\quad
32{,}718
\end{equation}
normalization SVD calls, respectively. The final effective spectral norms differ from unity by at most $1.9\times10^{-8}$.

These counts characterize the exact-SVD matched control used to enforce
$\|W\|_2=1$ numerically. They should not be interpreted as claiming that exact SVD is the only classical method for norm control; approximate power iteration and structurally orthogonal classical parametrizations are alternatives.

\subsection{Approximate Spectral-Normalization Stress Test}
\label{sup:approx_sn}

To isolate spectral-scale estimation error, the six matched-dense checkpoints are frozen and their exact normalization scales are replaced post hoc by cold-start power-iteration estimates with
\begin{equation}
k\in\{1,2,4,8,16,32,64\}.
\end{equation}
Sixteen predetermined estimator initializations are used, giving $4032$ individual matrix estimates and $672$ complete approximate-normalization realizations.

\begin{table*}[t]
\centering
\caption{Frozen-checkpoint approximate spectral-normalization study. ``Norm err.'' denotes median absolute relative spectral-norm estimation error.}
\label{sup:tab:approx_sn}
\small
\setlength{\tabcolsep}{3pt}
\begin{tabular*}{\textwidth}{@{\extracolsep{\fill}}l c c c c c c}
\hline
PDE & $k$ & Norm err. & Med. $\max\tau$ & Min. margin & Med. $\widetilde\chi_Z/\chi_Z$ & P95 ratio \\
\hline
Heat & 1 & $0.1775$ & $1.4016$ & $0.3538$ & $1.8792$ & $3.7202$ \\
Heat & 8 & $0.0210$ & $1.0817$ & $0.5699$ & $1.0912$ & $1.1691$ \\
Heat & 16 & $1.38\times10^{-3}$ & $1.0235$ & $0.6054$ & $1.0165$ & $1.0537$ \\
Heat & 32 & $1.24\times10^{-5}$ & $1.0018$ & $0.6060$ & $1.0016$ & $1.0204$ \\
Burgers & 1 & $0.2433$ & $1.4643$ & $0.0099$ & $2.9397$ & $135.05$ \\
Burgers & 8 & $3.18\times10^{-3}$ & $1.0485$ & $0.4703$ & $1.0321$ & $1.3272$ \\
Burgers & 16 & $1.57\times10^{-5}$ & $1.0027$ & $0.5067$ & $1.00005$ & $1.0138$ \\
Burgers & 32 & $1.16\times10^{-9}$ & $1.0002$ & $0.5067$ & $1.00000$ & $1.00001$ \\
\hline
\end{tabular*}
\end{table*}

Every tested approximate model retains a positive residual contraction margin and all sampled representations satisfy the corresponding propagated sufficient bounds. The experiment therefore isolates sensitivity of the certificate to spectral-scale estimation rather than retraining adaptation.

\section{Contextual In-Domain Uncertainty Baselines}
\label{sup:context_uq}

For context, we report the archived in-domain MC Dropout \cite{gal2016dropout}, Deep Ensemble \cite{lakshminarayanan2017simple}, and SWAG \cite{maddox2019simple} experiments from the earlier study. These runs were conducted under the earlier direct in-domain UQ protocol and do \emph{not} share the frozen Heat-to-Burgers transfer protocol, matched architecture, or model-selection procedure used for the primary QO--dense comparison in the current paper. They are therefore reported as contextual baselines only and are not used for matched pairwise claims.

\begin{table*}[t]
\centering
\caption{Archived in-domain UQ baselines from the earlier experimental protocol. These results are contextual and are not directly matched to the frozen Heat-to-Burgers transfer experiment in the main paper.}
\label{sup:tab:archived_uq}
\small
\setlength{\tabcolsep}{3pt}
\begin{tabular*}{\textwidth}{@{\extracolsep{\fill}}l l c c c c c c}
\hline
PDE & Method & RMSE & MNLL & PICP2 & MPIW3 & EAC & DAC \\
\hline
Heat & MC Dropout & $1.18\times10^{-2}$ & $-2.35$ & $1.0000$ & $0.2570$ & $0.4766$ & $0.1528$ \\
Heat & Deep Ensemble & $1.26\times10^{-4}$ & $-7.47$ & $0.9034$ & $0.0007$ & $0.2461$ & $0.1181$ \\
Heat & SWAG & $1.34\times10^{-3}$ & $-5.04$ & $1.0000$ & $0.0131$ & $0.2569$ & $-0.0897$ \\
\hline
Burgers & MC Dropout & $5.52\times10^{-2}$ & $-1.57$ & $0.8731$ & $0.2584$ & $0.2877$ & $0.2034$ \\
Burgers & Deep Ensemble & $5.49\times10^{-2}$ & $-2.44$ & $1.0000$ & $0.2628$ & $0.9360$ & $0.0352$ \\
Burgers & SWAG & $1.44\times10^{-2}$ & $9.31$ & $0.5141$ & $0.0154$ & $0.2622$ & $0.0405$ \\
\hline
\end{tabular*}
\end{table*}

The archived comparisons show why the present manuscript separates absolute calibration, error awareness, and distance awareness rather than treating any single uncertainty metric as sufficient. However, because the protocols differ, the current matched QO--dense transfer experiment remains the appropriate control for isolating the effect of structural orthogonality.

\section{Additional Covariance Sensitivity}
\label{sup:covariance_sensitivity}

To isolate uncertainty scale from representation learning, every Burgers checkpoint, predictive mean, latent representation, random-feature realization, and supervised design matrix is held fixed while only $(\lambda,\sigma_y^2)$ is varied.

At fixed $\sigma_y^2=3\times10^{-8}$, reducing $\lambda$ from $10^8$ to $10^2$ increases the mean predictive standard deviation from approximately
\begin{equation}
1.93\times10^{-4}
\quad\text{to}\quad
6.69\times10^{-2}
\end{equation}
for QO and from
\begin{equation}
1.95\times10^{-4}
\quad\text{to}\quad
7.33\times10^{-2}
\end{equation}
for dense. The corresponding PICP2 values reach $0.783$ and $0.872$, respectively.

At fixed $\lambda=10^8$, increasing $\sigma_y^2$ to $10^{-2}$ produces predictive standard deviations of approximately $0.1$ and PICP2 values of $0.916$ for QO and $0.968$ for dense.

These changes occur with the deterministic representation held fixed, directly illustrating that the bi-Lipschitz geometry does not determine the absolute covariance scale.

\section{Additional Seed-Level Results}
\label{sup:seed_results}

All values in this section are taken directly from the archived per-seed outputs rather than reconstructed from aggregate means and standard deviations. Differences are QO minus dense unless stated otherwise.

\begin{table*}[t]
\centering
\caption{Held-out midpoint-grid forward QO-FiLM versus matched dense-FiLM results by seed. Here $\Delta=\mathrm{QO}-\mathrm{Dense}$; negative differences favor QO because both metrics are errors.}
\label{sup:tab:forward_seed}
\setlength{\tabcolsep}{4pt}
\resizebox{\textwidth}{!}{%
\begin{tabular}{l c c c c c c c}
\hline
PDE & Seed & QO Rel. $L_2$ & Dense Rel. $L_2$ & $\Delta$ Rel. $L_2$ & QO Max & Dense Max & $\Delta$ Max \\
\hline
Burgers & 42 & $4.8248091\times10^{-4}$ & $4.9272887\times10^{-4}$ & $-1.0247964\times10^{-5}$ & $1.0342820\times10^{-3}$ & $1.2351376\times10^{-3}$ & $-2.0085559\times10^{-4}$ \\
Burgers & 123 & $7.2822955\times10^{-4}$ & $3.9844873\times10^{-4}$ & $3.2978082\times10^{-4}$ & $1.8571420\times10^{-3}$ & $1.0771743\times10^{-3}$ & $7.7996770\times10^{-4}$ \\
Burgers & 456 & $3.4708877\times10^{-4}$ & $4.0227470\times10^{-4}$ & $-5.5185928\times10^{-5}$ & $9.2034876\times10^{-4}$ & $1.0963706\times10^{-3}$ & $-1.7602179\times10^{-4}$ \\
Adv.--Diff. 1D & 42 & $3.6673588\times10^{-3}$ & $1.6205814\times10^{-3}$ & $2.0467774\times10^{-3}$ & $8.0929041\times10^{-2}$ & $5.7414544\times10^{-2}$ & $2.3514497\times10^{-2}$ \\
Adv.--Diff. 1D & 123 & $5.5022091\times10^{-3}$ & $4.6525932\times10^{-3}$ & $8.4961585\times10^{-4}$ & $1.7145108\times10^{-1}$ & $1.2503805\times10^{-1}$ & $4.6413033\times10^{-2}$ \\
Adv.--Diff. 1D & 456 & $4.5587178\times10^{-3}$ & $2.8259818\times10^{-3}$ & $1.7327360\times10^{-3}$ & $8.8899882\times10^{-2}$ & $6.7002223\times10^{-2}$ & $2.1897659\times10^{-2}$ \\
Adv.--Diff. 2D & 42 & $3.1396022\times10^{-3}$ & $4.4881380\times10^{-3}$ & $-1.3485357\times10^{-3}$ & $0.39316825$ & $0.44160451$ & $-0.048436258$ \\
Adv.--Diff. 2D & 123 & $3.6698076\times10^{-3}$ & $4.8889471\times10^{-3}$ & $-1.2191395\times10^{-3}$ & $0.38488355$ & $0.47652906$ & $-0.091645504$ \\
Adv.--Diff. 2D & 456 & $2.4219410\times10^{-3}$ & $3.3852193\times10^{-3}$ & $-9.6327837\times10^{-4}$ & $0.22517691$ & $0.32461284$ & $-0.099435927$ \\
Adv.--Diff. 3D & 42 & $0.024354317$ & $0.022355938$ & $0.0019983781$ & $5.9511365$ & $4.5924287$ & $1.3587078$ \\
Adv.--Diff. 3D & 123 & $0.019694427$ & $0.020719738$ & $-0.0010253101$ & $4.2303758$ & $5.5346208$ & $-1.3042450$ \\
Adv.--Diff. 3D & 456 & $0.023206001$ & $0.024852629$ & $-0.0016466280$ & $6.4453763$ & $6.4293351$ & $0.016041129$ \\
\hline
\end{tabular}}
\end{table*}

\begin{table}[t]
\centering
\caption{Seed-level paired UQ and geometry differences, QO minus dense. Negative $\Delta$RMSE, $\Delta$CE, and $\Delta\chi_Z$ favor QO; positive $\Delta$DAC favors QO.}
\label{sup:tab:uq_paired}
\setlength{\tabcolsep}{4pt}
\resizebox{\columnwidth}{!}{%
\begin{tabular}{l c c c c c}
\hline
PDE & Seed & $\Delta$RMSE & $\Delta$CE & $\Delta$DAC & $\Delta\chi_Z$ \\
\hline
Heat & 42 & $6.0867871\times10^{-5}$ & $0.061730124$ & $-0.029115353$ & $1.4872786$ \\
Heat & 123 & $7.3416102\times10^{-5}$ & $0.084207431$ & $-0.056330855$ & $-3.6601670$ \\
Heat & 456 & $7.7002152\times10^{-5}$ & $0.045849328$ & $-0.023016555$ & $-2.3189660$ \\
Burgers & 42 & $0.037456382$ & $0.0039211842$ & $0.0044476751$ & $-8.1705707$ \\
Burgers & 123 & $-0.012534511$ & $-0.0034310362$ & $0.034778491$ & $-8.6697554$ \\
Burgers & 456 & $0.020775182$ & $0.0056857171$ & $0.017217968$ & $-6.3417597$ \\
\hline
\end{tabular}}
\end{table}

\begin{table*}[t]
\centering
\caption{Held-out midpoint-grid amplitude ablation for the six Burgers/2D advection--diffusion seed pairs for which restoration-only runs were completed. ``Direction'' denotes direction-only RBS, and ``Restore'' denotes exact norm restoration without FiLM.}
\label{sup:tab:amplitude_ablation}
\setlength{\tabcolsep}{3.5pt}
\resizebox{\textwidth}{!}{%
\begin{tabular}{l c c c c c c c c c c}
\hline
PDE & Seed & Dir. $L_2$ & Restore $L_2$ & QO-FiLM $L_2$ & Dense-FiLM $L_2$ & Dir. Max & Restore Max & QO-FiLM Max & Dense-FiLM Max \\
\hline
Burgers & 42 & $0.0012335188$ & $0.011502082$ & $0.00048248091$ & $0.00049272887$ & $0.0035865056$ & $0.017772485$ & $0.0010342820$ & $0.0012351376$ \\
Burgers & 123 & $0.00095470964$ & $0.0076034826$ & $0.00072822955$ & $0.00039844873$ & $0.0031563799$ & $0.011405635$ & $0.0018571420$ & $0.0010771743$ \\
Burgers & 456 & $0.0010316402$ & $0.0090204486$ & $0.00034708877$ & $0.00040227470$ & $0.0034039944$ & $0.013125439$ & $0.00092034876$ & $0.0010963706$ \\
Adv.--Diff. 2D & 42 & $0.0031151921$ & $0.020873279$ & $0.0031396022$ & $0.0044881380$ & $0.29811336$ & $3.2186403$ & $0.39316825$ & $0.44160451$ \\
Adv.--Diff. 2D & 123 & $0.0030570414$ & $0.025539515$ & $0.0036698076$ & $0.0048889471$ & $0.37542432$ & $3.8427937$ & $0.38488355$ & $0.47652906$ \\
Adv.--Diff. 2D & 456 & $0.0022019604$ & $0.016732260$ & $0.0024219410$ & $0.0033852193$ & $0.18132422$ & $3.5461603$ & $0.22517691$ & $0.32461284$ \\
\hline
\end{tabular}}
\end{table*}

Restoration-only is worse than direction-only in all six completed Burgers/2D advection--diffusion pairs for both relative $L_2$ and maximum error, while QO-FiLM improves over restoration-only in all six pairs for both metrics. Restoration-only runs were not performed for one- or three-dimensional advection--diffusion and no values are reconstructed or imputed for those cases.

\begin{table}[t]
\centering
\caption{Per-seed finite-readout certificate constants for the frozen QO checkpoints. The sufficient tolerances use $\zeta=0.5$ and are computed independently for each seed before aggregation.}
\label{sup:tab:finite_readout_seed}
\resizebox{\columnwidth}{!}{%
\begin{tabular}{l c c c c c}
\hline
PDE & Seed & $C_Z$ & $r_{\rm read}/\epsilon$ & $\epsilon_{\max}(0.05)$ & $\epsilon_{\max}(0.10)$ \\
\hline
Heat & 42 & $27.98483245$ & $173.5668274$ & $1.4403674\times10^{-4}$ & $2.8807348\times10^{-4}$ \\
Heat & 123 & $26.39449006$ & $156.6532734$ & $1.5958811\times10^{-4}$ & $3.1917622\times10^{-4}$ \\
Heat & 456 & $27.03672924$ & $157.0550314$ & $1.5917987\times10^{-4}$ & $3.1835975\times10^{-4}$ \\
Burgers & 42 & $36.58640008$ & $269.5065991$ & $9.2762107\times10^{-5}$ & $1.8552421\times10^{-4}$ \\
Burgers & 123 & $50.40449000$ & $592.4780371$ & $4.2195657\times10^{-5}$ & $8.4391314\times10^{-5}$ \\
Burgers & 456 & $40.15973730$ & $332.3997900$ & $7.5210637\times10^{-5}$ & $1.5042127\times10^{-4}$ \\
\hline
\end{tabular}}
\end{table}

The paired design is particularly important because each QO--dense pair shares data, validation points, non-map initialization, random-feature realization, and optimization protocol. The nonlinear derived quantities in Table~\ref{sup:tab:finite_readout_seed} are computed per seed before taking the mean and population standard deviation reported elsewhere.

\section{Reproducibility and Interpretation Notes}
\label{sup:interpretation}

The resource quantities used throughout the paper distinguish:
(i) exact classical simulation work,
(ii) sequential quantum circuit depth,
(iii) total RBS gate count,
(iv) signed-readout access-model repetitions, and
(v) dense spectral-normalization work.

All PDE predictions reported in the paper are generated by classical simulation of the learned QO maps. Consequently, classical-simulator wall-clock time is not presented as a proxy for quantum hardware runtime. The computational claims concern the stated access model for executing a learned orthogonal transformation and do not establish an end-to-end quantum-training speedup.

The contextual MC Dropout, Deep Ensemble, and SWAG results in Section~\ref{sup:context_uq} do not share the model-selection, transfer, or architectural-control protocol of the matched QO--dense experiment. Their purpose is to retain comparison with standard UQ approaches without conflating unmatched historical experiments with the decisive architectural control.

The finite-readout perturbation experiment validates the bounded directional-error event assumed in the theoretical analysis and is not itself a simulation of a tomography protocol. Separately, Section~\ref{sup:readout_scale} reports explicit state-preparation/measurement repetition counts for the signed $\ell_\infty$ tomography construction of \cite{kerenidis2019quantum}. These construction-level counts do not model device-specific measurement latency, hardware parallelism, decoherence, gate infidelity, error correction, or the multiplicity of an end-to-end training workload. Establishing the practical classical--quantum crossover therefore remains an empirical hardware question.

\bibliographystyle{IEEEtran}
\bibliography{orthobib}